\documentclass[acmsmall]{acmart}
\AtBeginDocument{%
  }

\setcopyright{none}     
\makeatletter
\renewcommand\@formatdoi[1]{\ignorespaces} 
\makeatother

\usepackage{gensymb}
\usepackage{subfigure}
\usepackage{array}
\usepackage{soul}
\usepackage[normalem]{ulem}
\usepackage{algorithm}
\usepackage{algpseudocode}
\usepackage{xcolor}
\usepackage{multirow}

\begin{document}

\title{Viewport-based Neural 360\degree~Image Compression}


\author{Jingwei Liao}
\affiliation{%
  \institution{George Mason University}
  \city{Fairfax}
  \country{USA}}
\email{jliao2@gmu.edu}

\author{Bo Chen}
\affiliation{%
  \institution{University of Illinois Urbana-Champaign}
  \city{Urbana}
  \country{USA}}
\email{boc2@illinois.edu}

\author{Klara Nahrstedt}
\affiliation{%
  \institution{University of Illinois Urbana-Champaign}
  \city{Urbana}
  \country{USA}}
\email{klara@illinois.edu}

\author{Zhisheng Yan}
\affiliation{%
  \institution{George Mason University}
  \city{Fairfax}
  \country{USA}}
\email{zyan4@gmu.edu}

\begin{abstract}
Given the popularity of 360\degree~images on social media platforms, 360\degree~image compression becomes a critical technology for media storage and transmission. Conventional 360\degree~image compression pipeline projects the spherical image into a single 2D plane, leading to issues of oversampling and distortion. In this paper, we propose a novel viewport-based neural compression pipeline for 360\degree~images. By replacing the image projection in conventional 360\degree~image compression pipelines with viewport extraction and efficiently compressing multiple viewports, the proposed pipeline minimizes the inherent oversampling and distortion issues.
However, viewport extraction impedes information sharing between multiple viewports during compression, causing the loss of global information about the spherical image.
To tackle this global information loss, we design a 
neural viewport codec to capture global prior information across multiple viewports and maximally compress the 
viewport data. The viewport codec is empowered by a transformer-based ViewPort ConText (VPCT) module that
can be integrated with canonical learning-based 2D image compression structures. 
We compare the proposed pipeline with existing 360\degree~image compression models and conventional 360\degree~image compression pipelines building on learning-based 2D image codecs and standard hand-crafted codecs.  
Results show that our pipeline saves an average of $14.01\%$ bit consumption compared to the best-performing 360\degree~image compression methods without compromising quality. The proposed VPCT-based codec also outperforms existing 2D image codecs in the viewport-based neural compression pipeline. Our code can be found at: \href{https://github.com/Jingwei-Liao/VPCT}{https://github.com/Jingwei-Liao/VPCT}.
\end{abstract}




\maketitle

\section{Introduction}
The rise of affordable panoramic cameras~\cite{panoramic_camera2} and head-mounted displays~\cite{hmd1} has made 360\degree~images accessible on the majority of social media platforms, including Facebook and YouTube. These high-resolution 360\degree~images provide immersive experiences to everyday users, but necessitate increased storage space and network bandwidth. To minimize storage space and bandwidth consumption, 360\degree~image compression~\cite{structure_map,oslo,pseudocylindrical,360video1,360video2,360video3,360video4, jll} has played a pivotal role in today's social media platforms. 
\begin{figure}[htbp]
  \centering
  \includegraphics[width=0.75\linewidth]{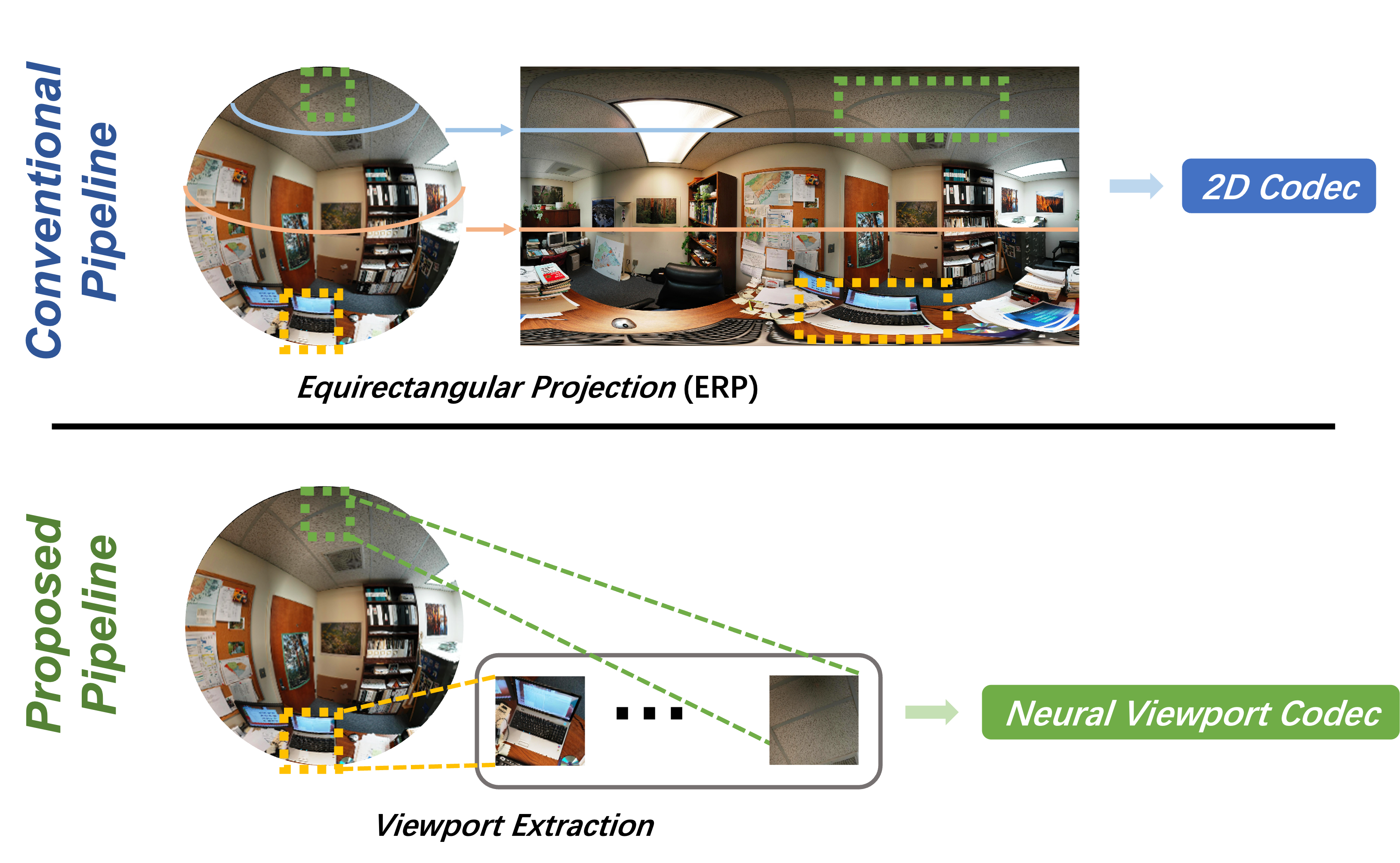}
  \caption{360\degree~image compression pipeline.
  \label{fig:normal_pipeline}}
\end{figure}

As depicted in Fig.~\ref{fig:normal_pipeline} (top), the conventional 360\degree~image compression pipeline consists of two steps~\cite{360overview}. The pre-processing step first maps spherical image content onto a 2D plane through \textit{image projection} such as the widely used equirectangular projection (ERP)~\cite{360overview, vpextract}. Subsequently, the acquired 2D content is compressed using 2D image compression. The 2D image compression has been achieved by either a hand-crafted codec in early works, e.g., JPEG and JPEG2000, or more recently a learning-based neural codec that learns a minimal representation of image pixels for enhancing compression efficiency~\cite{factorized, hyperprior, joint, tsai2018learning, gaussian_mixture, reference, hu2020coarse, liu2020learned, jiang2023multi,chen2022context}.

Despite the consensus about the two-step 360\degree~image compression~\cite{pseudocylindrical,structure_map,oslo}, we observe that image projection results in two inherent issues that impact compression performance -- \textit{oversampling} and \textit{distortion}. Oversampling arises from the fact that the spherical image's radius in high-latitude areas is shorter than that in low-latitude areas. However, image projection methods like ERP project the content of different latitude areas onto a 2D image with a uniform length, thereby introducing extraneous pixels into the projected image. As illustrated in Fig.~\ref{fig:normal_pipeline} (top), the orange and blue curves mark the image content located at different latitudes of the spherical image, but the blue curve is oversampled to the same length as the orange line in the 2D ERP image. On the other hand, distortion is the alteration of the relative geometric positions between the pixels after incorporating these extraneous pixels. For example, straight lines may be warped into curves, as marked by the green and yellow boxes in Fig.~\ref{fig:normal_pipeline} (top). 
As a result, oversampling introduces additional interpolation pixels for compression, increasing bit consumption, while distortion can alter the original geometric positional relationship between pixels, potentially reducing reconstruction quality, both impairing the performance of 360\degree~image compression.

While several learning-based neural 360\degree~images codecs have been recently designed to mitigate the oversampling and distortion issues~\cite{structure_map,pseudocylindrical}, they still utilize the same conventional pipeline.  
Since a 360\degree~image still has to undergo image projection before being compressed, these codecs inevitably introduce oversampling and distortion, affecting the compression performance.

To bridge this gap, we propose a novel viewport-based neural compression pieplien for 360\degree~images. As shown in Fig.~\ref{fig:normal_pipeline} (bottom), our pipeline includes a viewport extraction step that replaces the image projection in the conventional pipeline and a neural viewport codec that compresses the extracted viewports. In contrast to image projection which obtains a single 2D image by continuously projecting the entire spherical image, viewport extraction uses multiple discrete 2D viewports to cover the spherical image. Since a spherical area to be extracted as a viewport is sufficiently small and can be approximated as a 2D plane, oversampling barely occurs when mapping this spherical area onto the viewport. Given that few spherical pixels are oversampled, the geometric positions between most pixels remain consistent on the viewports, thereby minimizing the distortion of the pre-processed 2D content to be compressed. As demonstrated in Fig.~\ref{fig:normal_pipeline}, the two spherical areas marked by the green and yellow boxes introduce significantly less geometric distortion when displayed on the viewport than on the ERP. Similarly, the same content is represented by fewer pixels in the viewport than in the ERP.

Realizing the viewport-based neural 360\degree~image compression requires us to overcome new challenges unaddressed by prior research. Specifically, the rectangular viewports cannot evenly cover the entire sphere, causing content overlaps between extracted viewports and potentially increasing bit consumption. Since existing handcrafted and neural 2D image codecs are all designed to compress content within a single image, they would individually compress each extracted viewport in our proposed pipeline. The image compression model cannot explore inter-pixel dependencies between viewports, which could degrade the compression performance. In essence, these challenges stem from the fact that global prior information of the spherical image is lost during viewport extraction. 
Therefore, we propose the neural viewport codec in our pipeline that integrates the global prior information into the entropy model of learning-based compression through the ViewPort ConText (VPCT) module. VPCT is empowered by an intra-view block using a customized self-attention layer to capture causal contextual information within regions of a viewport, as well as an inter-view block employing a novel cross-view attention layer to capture correlations between different viewports. 

The benefits of the viewport-based pipeline are two-fold. First, compared with previous 2D codecs that can only process pixels or latent representations within a single image, our VPCT-enhanced codec can capture global prior information from multiple viewports. Consequently, we can achieve the desired viewport compression performance while minimizing the oversampling and distortion issues. Second, by integrating the VPCT module, we can seamlessly introduce global prior information into canonical entropy models without modifying their structure. This ensures compatibility with existing learning-based 2D compression models and their respective compression acceleration methods. 

We compare the viewport-based neural compression pipeline with three categories of baselines including specifically designed 360\degree~image compression models as well as conventional 360\degree~image compression pipelines building on learning-based 2D codecs and standard hand-crafted codecs. We also compare the neural viewport codec with state-of-the-art (SOTA) learning-based 2D codecs and hand-crafted 2D codecs in the viewport-based pipeline. The proposed work outperforms the existing state-of-the-art (SOTA) method in average bit savings by a significant margin of $14.01\%$, without compromising quality. Our results also show that the viewport-based pipeline is compatible with various canonical 2D compression structures and achieves consistent gains. The contributions of this paper are as follows.

\begin{itemize}
    \item We propose the first viewport-based 360\degree~image compression pipeline to minimize the oversampling and distortion issues.
    \item We design the VPCT module, compatible with existing learning-based 2D image codecs, to handle the global information loss introduced by viewport extraction. 
    \item The proposed pipeline achieves SOTA performance on 360\degree~image compression, outperforming previous methods.
\end{itemize}


\section{Related Work}

\subsection{Learning-based 2D Image Compression}
Compared with traditional hand-crafted image compression standards, learning-based neural compression has recently achieved outstanding performance through various deep learning techniques, including auto-encoders~\cite{factorized,hyperprior,joint,chen2021deep}, recurrent neural networks~\cite{variable-rate2}, generative adversarial networks~\cite{gan3}, and transformers~\cite{entroformer}.
These compression methods first transform the input image into a latent representation by an encoder. Subsequently, 
their entropy model compresses the latent representation by estimating its marginal distribution. Finally, the compressed latent representation is losslessly converted into a bitstream using entropy coding, such as Huffman coding. To reconstruct the image, the bitstream undergoes a set of reverse operations. 
Enhancing the estimation accuracy of the entropy model has become the primary approach to improving compression performance because the estimation determines the bitstream length \cite{context_adaptive,reference}. However, existing entropy models can only capture local prior information within a single image, rendering them incapable of handling redundant information between different viewports. Thus, we propose the VPCT module to capture the global prior information of a spherical image across viewports.
    

\subsection{Learning-based 360\degree~Image Compression}
Since directly applying 2D compression models to spherical images is infeasible, most 360\degree~image compression methods project spherical images onto a 2D planar format~\cite{360overview}. However, common image projection techniques, such as ERP, introduce oversampling and distortion that adversely impact compression performance. To mitigate oversampling, a structure map was incorporated into the 2D compression model, allocating fewer bits to oversampled areas~\cite{structure_map}. Similarly, another model employed a greedy approach to identify the optimal configuration for image projection~\cite{pseudocylindrical}. Although these approaches alleviate oversampling to a certain degree, they fail to address the geometric distortion. 
The fundamental issue of these 360\degree~image compression models is that image projection is always performed, causing significant oversampling and distortion. 
In this paper, we propose a new pipeline that replaces image projection with viewport extraction to circumvent the issues.


\subsection{Viewport-based 360\degree~Video Streaming}
Viewport extraction has been widely used in 360\degree~video streaming systems to support viewport-adaptive viewing~\cite{vpc1,vpc2,vpc4,yi2020analysis,nguyen2018your, 360video}. During a video streaming session, these systems dynamically predict the region of a 360\degree~video that a user is likely to view in the upcoming period and then extract and stream the corresponding viewport to the user. Viewport-based 360\degree~video streaming significantly reduces the network bandwidth by only compressing and streaming a single viewport. However, it is important to note that 360\degree~video streaming is a different research problem from the 360\degree~image compression we consider in this paper. Rather than streaming \textit{a single viewport} to satisfy users' needs of on-the-fly viewing, our goal is to compress \textit{the entire 360\degree~spherical image} into a bitstream for future archive or transmission. Existing viewport-based 360\degree~video streaming systems typically re-use standardized hand-crafted codecs to compress the single viewport before streaming and do not explore the neural compression of multiple viewports or the whole sphere. 

\subsection{Transformer}
Self-attention techniques have been used in computer vision tasks~\cite{transformer_gan,transformer_object_detection, diffusion_data1, diffusion_data2, diffusion_data3, diffusion_data4, img_seg, quantum_com, shadow1, shadow2}. 
Recently, transformer models have been integrated into image compression~\cite{contextformer}. Lu et al.~\cite{tic} proposed the first transformer-based image compression model. EntroFormer~\cite{entroformer} utilized a transformer model to construct an entropy model capable of capturing longer dependencies in the latent representation compared to CNN-based entropy models. However, existing transformer-based compression focuses on discovering contextual relationships within a single image and does not address correlations between multiple images. 
We also note that recent cross-domain studies on adaptive filtering and pruning~\cite{unisdnet,llm_prune} are related to our design, as they improved efficiency by suppressing less informative components. However, they targeted video grounding and LLM pruning, respectively, rather than multi-viewport 360\degree~image compression. Hence, we design the transformer-based VPCT module to support the proposed neural viewport codec.

\section{Methodology}

\begin{figure}[!t]
  \centering
  \includegraphics[width=0.4\linewidth]{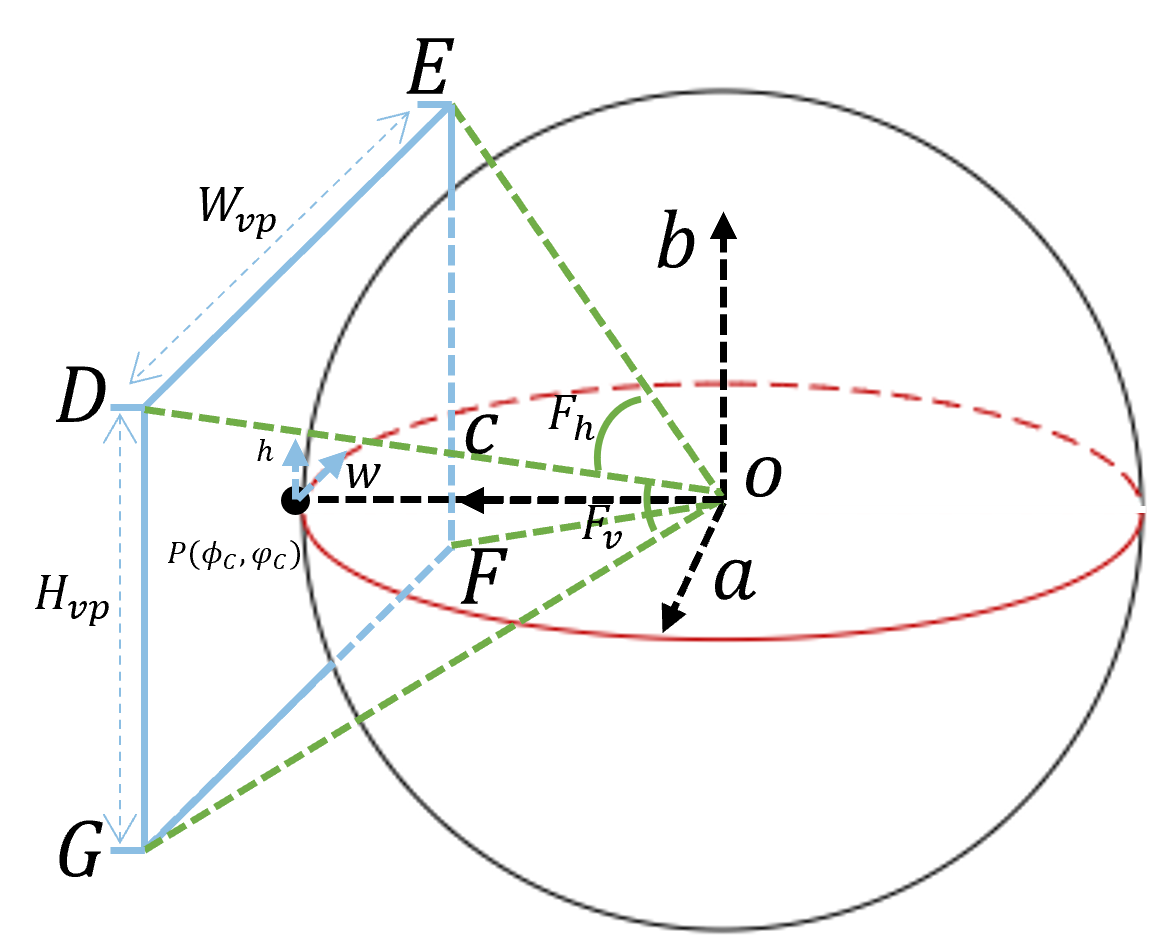}
  \caption{Viewport extraction process.}
  \label{fig:vp_ext}
\end{figure}
As shown in Fig.~\ref{fig:normal_pipeline} (bottom), the proposed viewport-based neural compression encompasses two steps. The initial step projects the spherical image onto various viewports through viewport extraction in order to minimize oversampling and distortion. The second step directs viewports to the neural viewport codec to maximally compress the data. 

\subsection{Viewport Extraction}
Fig.~\ref{fig:vp_ext} illustrates the process of extracting a viewport from the spherical image. The plane $DEFG$ denotes the viewport to be extracted. The center of the viewport plane $DEFG$ is tangent to the spherical image at point $P$.
Viewport extraction starts by corresponding the 2D coordinates ($w$, $h$) on the viewport with the Cartesian coordinates ($a$, $b$, $c$) on the spherical image. 
Then the pixel value in a specific position on the spherical image is placed into the corresponding position within the viewport. 
When the viewport is centered at 0\degree~latitude and 0\degree~longitude on a unit sphere (radius being $1.0$), the coordinate conversion between the viewport image and the spherical image is defined as,
\begin{center}
$
\begin{aligned}
a = \frac{ 2w }{W_{vp}}\tan\left ( F_h / 2\right),\ b = \frac{ 2h }{H_{vp}}\tan\left ( F_v / 2\right),\ c = 1.0
\end{aligned}
$
\end{center}
where $F_h$ and $F_v$ denote the horizontal Field of View (FoV) and vertical FoV of the viewport while $W_{vp}$ and $H_{vp}$ represent the width and height of the viewport, respectively.

By fixing the viewport window and then rotating the sphere, target areas of the spherical image can be extracted as additional viewports. 
The rotation matrix $R$ and the new spherical coordinates ($a'$, $b'$, $c'$) are defined as follows \cite{vpextract}.
\begin{gather*}
    R=\begin{bmatrix}
\cos \left(\varphi_{C}\right) & -\sin \left(\varphi_{C}\right) \sin \left(\phi_{C}\right) & \sin \left(\varphi_{C}\right) \cos \left(\phi_{C}\right) \\
0 & \cos \left(\phi_{C}\right) & \sin \left(\phi_{C}\right) \\
-\sin \left(\varphi_{C}\right) & -\cos \left(\varphi_{C}\right) \sin \left(\phi_{C}\right) & \cos \left(\varphi_{C}\right) \cos \left(\phi_{C}\right)
\end{bmatrix}\\
[a',b',c']^T = R [a,b,c]^T
\end{gather*}
where $\varphi_C$ and $\phi_C$ denote the longitude and latitude of the spherical position $P$, respectively. The values of various viewport centers are determined by the number of viewports extracted to cover the sphere, which is in turn determined by $F_h$ and $F_v$. Since small spherical areas can be approximated as 2D planes, the extracted viewports incur minimal oversampling and geometric distortion. 


\subsection{Neural Viewport Codec}

\begin{figure*}[]
    \centering
    \subfigure[General Structure]{
	\includegraphics[width=0.4\linewidth, height=0.25\textheight]{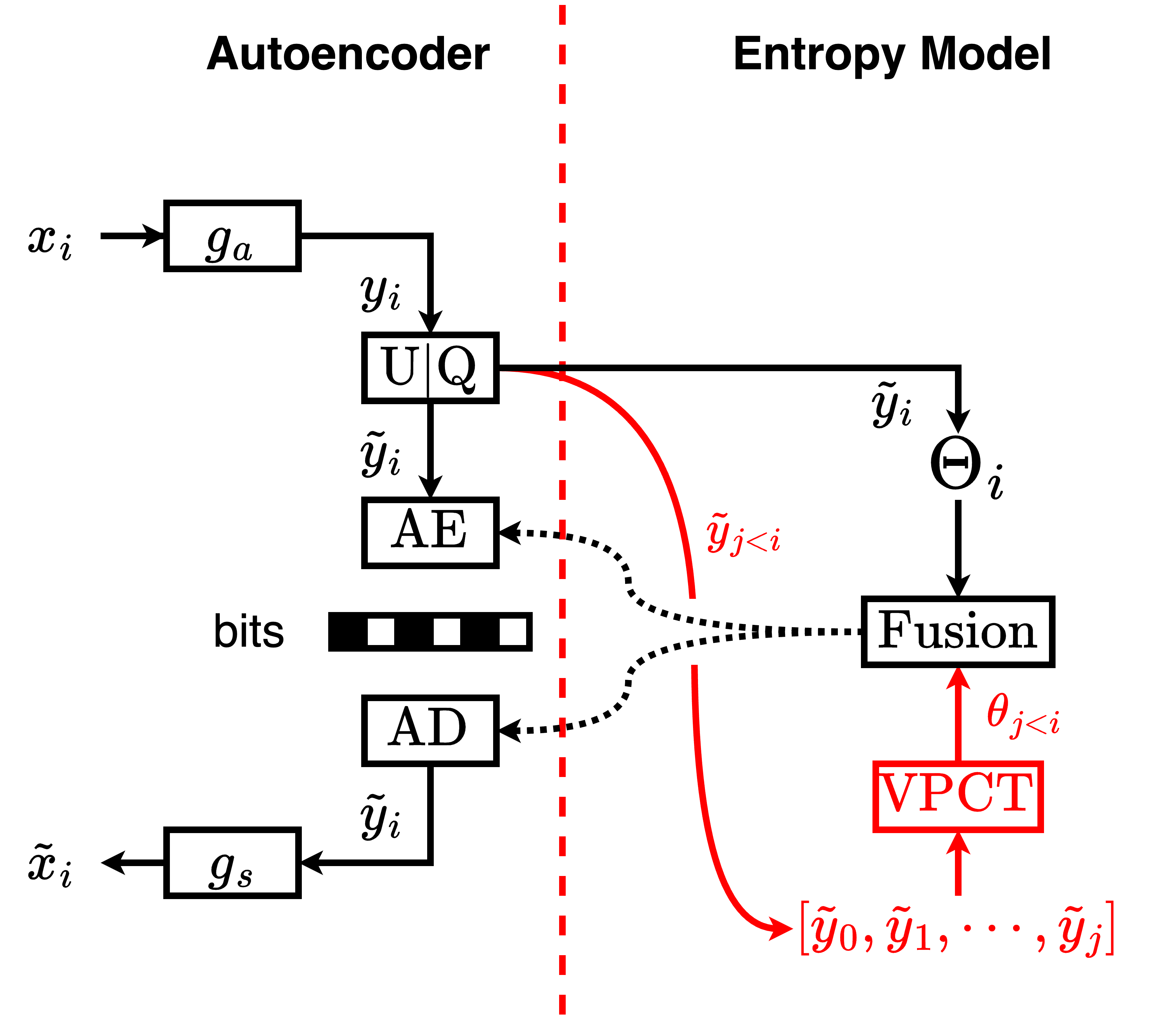}
        \label{fig:general_structure}
    }
    \subfigure[Factorized]{
	\includegraphics[width=0.4\linewidth, height=0.25\textheight]{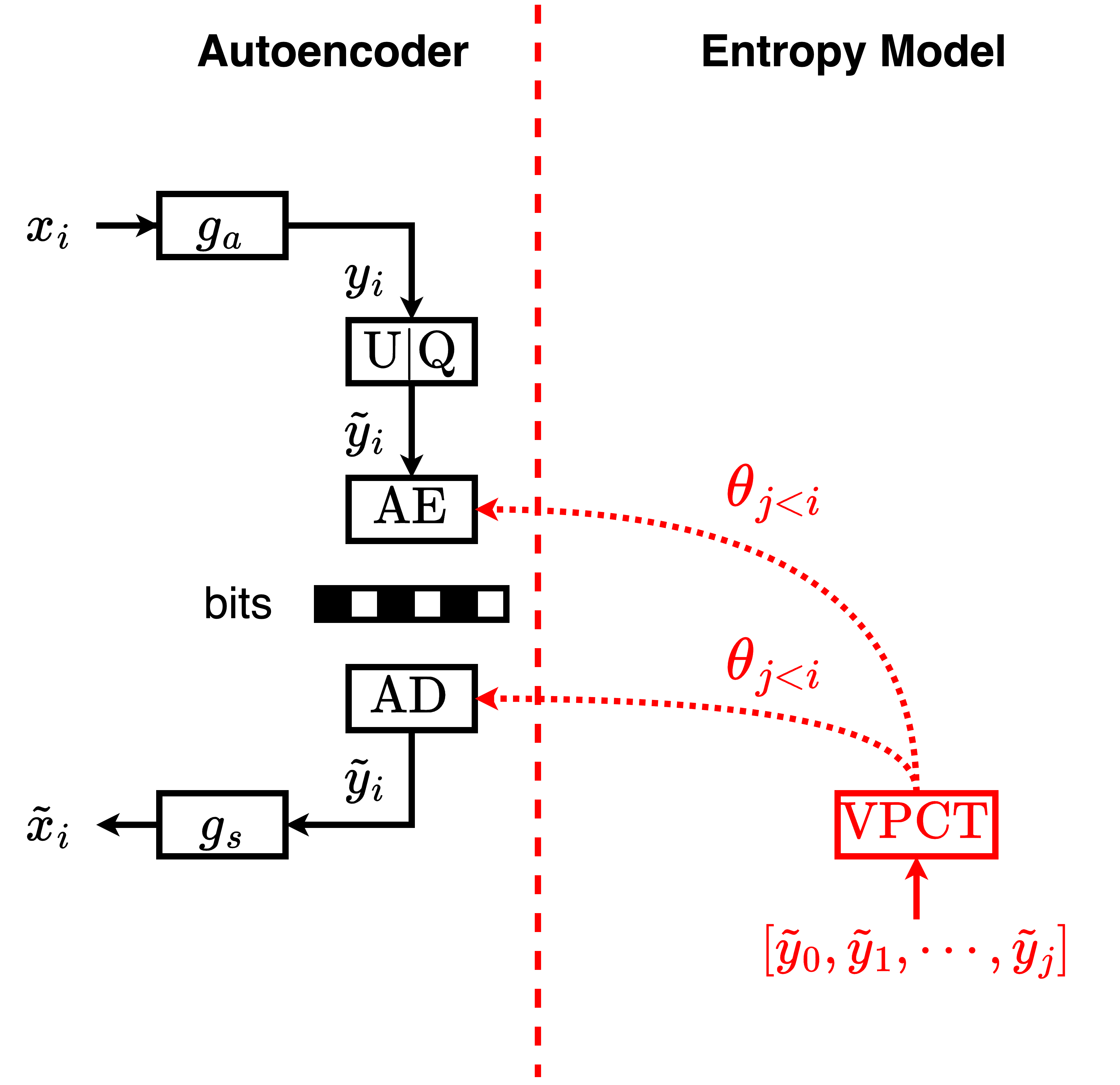}
        \label{fig:factorized}
    } \\
    \subfigure[Hyperprior]{
	\includegraphics[width=0.4\linewidth, height=0.25\textheight]{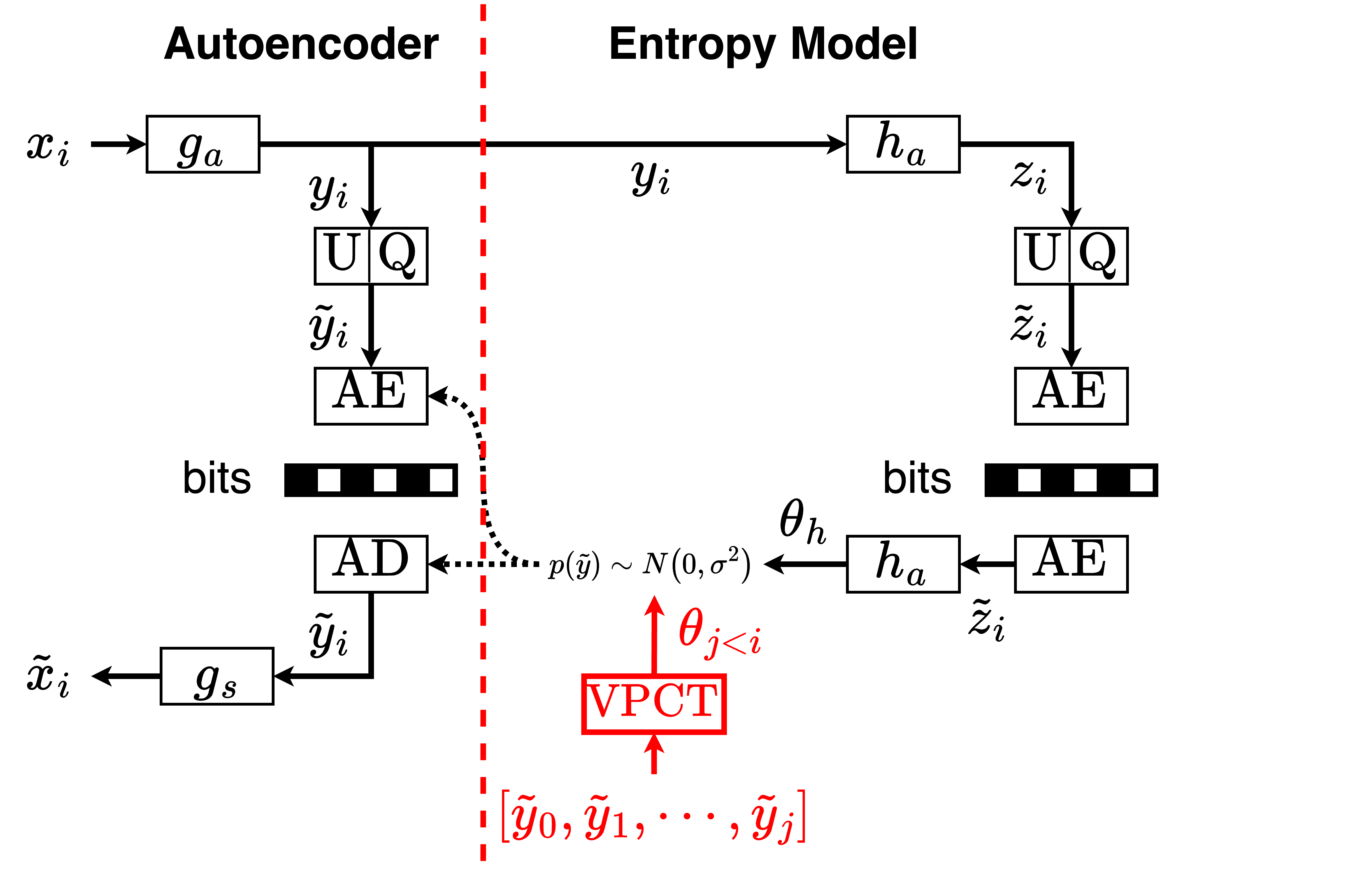}
        \label{fig:hyperprior}
    }
    \subfigure[Joint]{
	\includegraphics[width=0.4\linewidth, height=0.25\textheight]{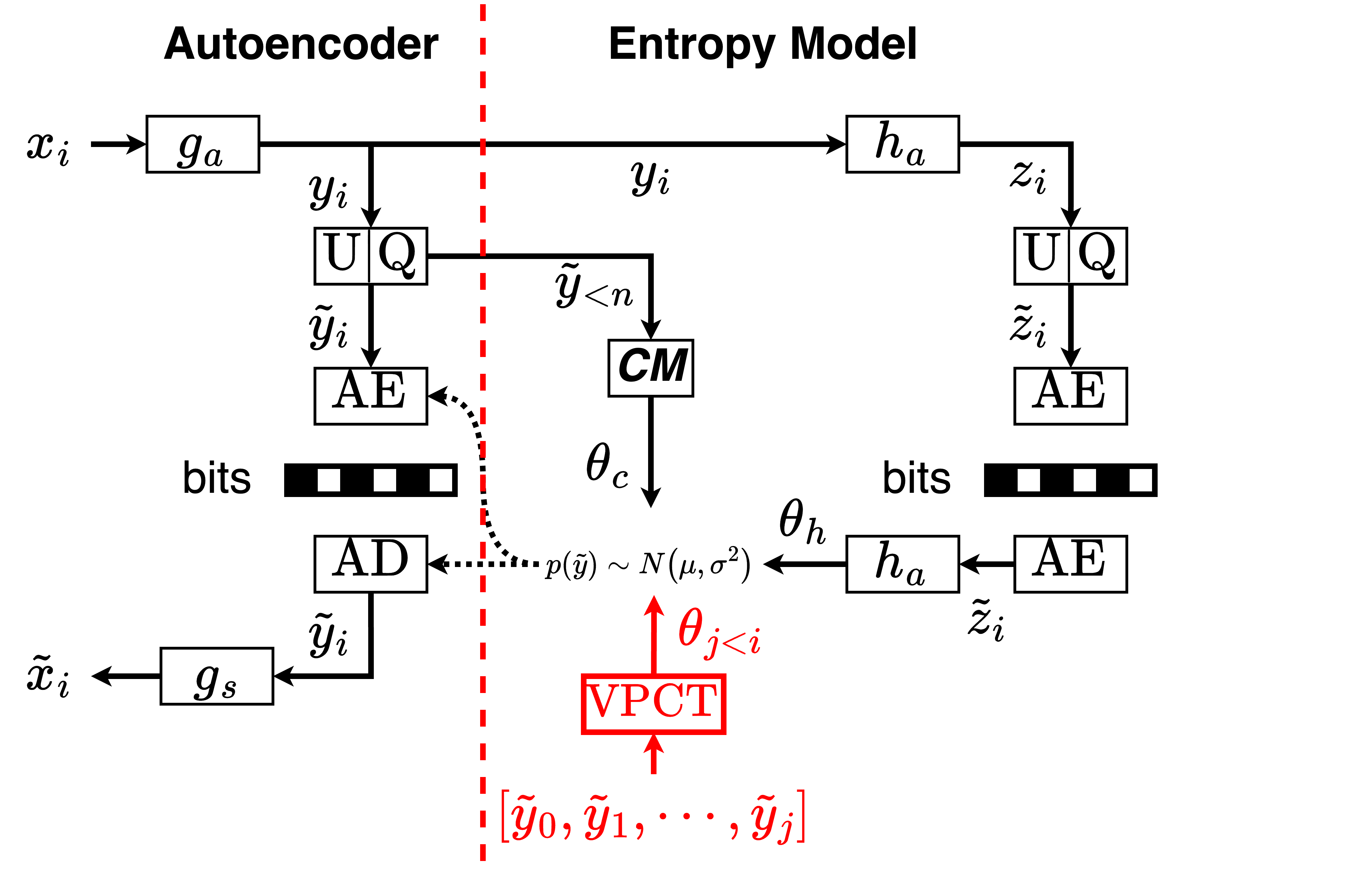}
        \label{fig:joint}
    } \\
    \subfigure[Reference]{
	\includegraphics[width=0.4\linewidth, height=0.25\textheight]{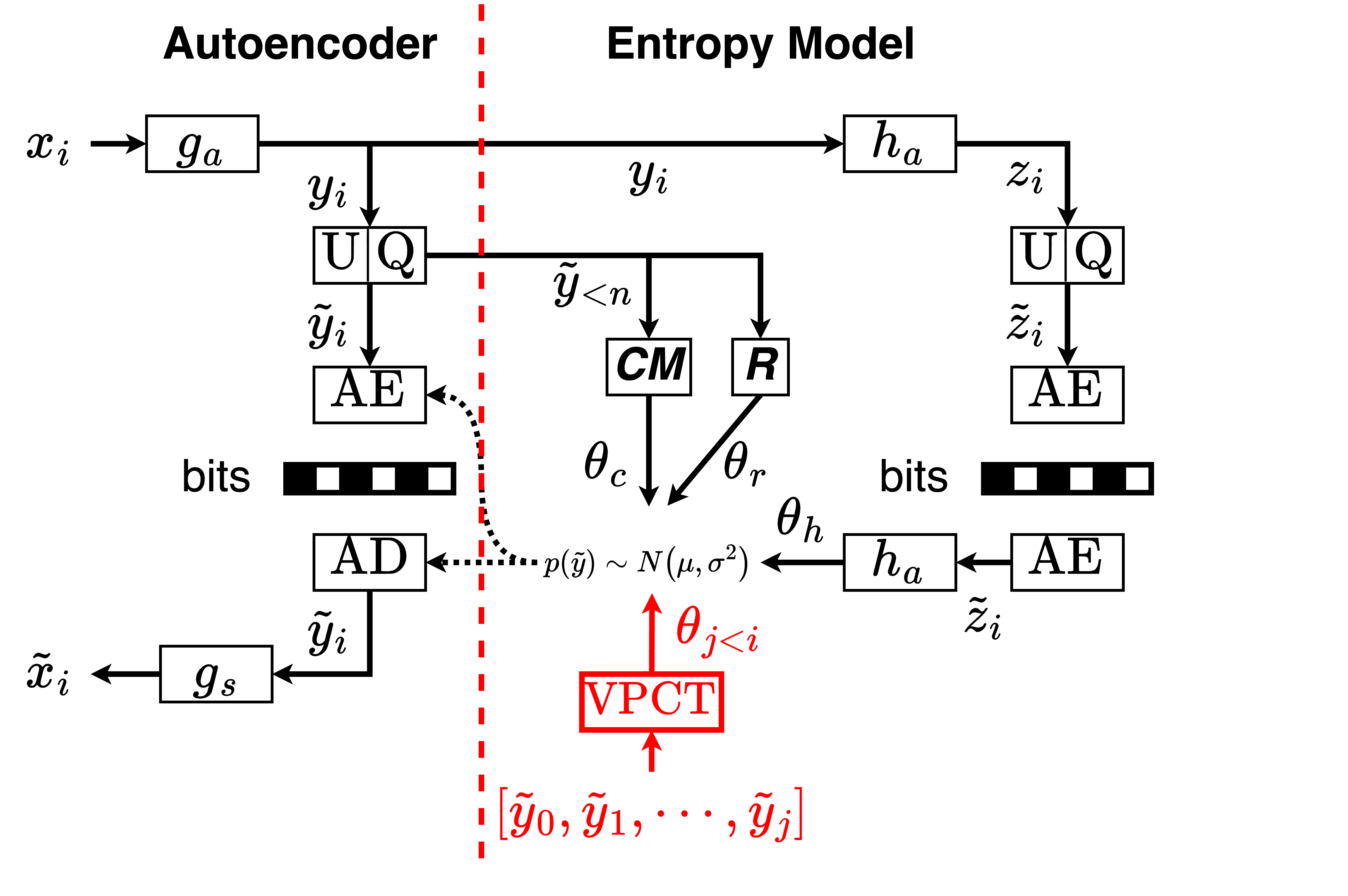}
        \label{fig:reference}
    } 
    \subfigure[Table of Notations]{
        \raisebox{25mm}{
        \scriptsize
        \begin{tabular}{cc}
        \toprule
        Notation & Description \\
        \midrule
        $x_i / \tilde{x}_i$& Raw/Reconstructed $i$-th input viewport image\\
        $y_i / \tilde{y}_i$& Raw/Quantized latent\\
        $z_i / \tilde{z}_i$& Raw/Quantized hyperprior latent\\
        $\tilde{y}_{<n}$ & Causal context of each element in $\tilde{y}_i$ \\
        $\tilde{y}_{j<i}$ & Previous viewports' quantized latents \\
        $g_a(\cdot) / g_s(\cdot)$& Encoder / Decoder\\
        $h_a(\cdot) / h_s(\cdot)$& Hyperprior Encoder / Decoder\\
        $\mathrm{U}|\mathrm{Q}$ & Uniform noise or round-based quantization \\
        AE/AD& Arithmetic encoding / decoding\\
        CM & Autoregressive context model \\
        R & Reference Model \\
        $\Theta_i$ & Local information \\
        $\theta_{j<i}$ & Cross-image information \\
        $\mathcal{N}(\mu, \sigma^2)$ & Gaussian distribution (Normal distribution) \\
        $\mu$ & Mean parameter of Gaussian \\
        $\sigma$ & Standard deviation parameter of Gaussian \\
        \bottomrule
        \end{tabular}
        \label{fig:parameters_table}
        }
    }
    \caption{Operational diagrams of learning-based compression models, along with how VPCT can be integrated. (a) general structure of the proposed viewport codec. (b-e) four canonical entropy models with the proposed VPCT. (f) table of notations.}
    \label{fig:entropy_models}
\end{figure*}

Fig.~\ref{fig:general_structure} overviews the proposed viewport codec, while the notations used in the following discussion are summarized in Fig.~\ref{fig:parameters_table}. The autoencoder extracts the latent representation $y_i$ from the viewport image $x_i$ through the encoder $g_{a}(\cdot)$, and reconstructs the viewport image $\tilde{x}_{i}$ from the discretized latent representation $\tilde{y}_{i}$ using the decoder $g_{s}(\cdot)$. The discretization from $y_i$ to $\tilde{y}_{i}$ is handled by the quantizer $\mathrm{U}|\mathrm{Q}$. During the training phase, the $\mathrm{U}|\mathrm{Q}$ is approximated by a uniform noise $\mathcal{U}(-\frac{1}{2}, \frac{1}{2})$ to enable gradient-based end-to-end training. During the inference phase, the $\mathrm{U}|\mathrm{Q}$ represents real round-based quantization. The discretized latent representation $\tilde{y}_i$ is then losslessly compressed using entropy coding methods, such as arithmetic coding, to generate a bitstream.

Entropy coding relies on a prior probability model to estimate the probability distribution of $\tilde{y}_i$, which is shared between the arithmetic encoder $\mathrm{AE}$ and arithmetic decoder $\mathrm{AD}$. This prior probability model is referred as the \textit{entropy} model.
Enhancing the entropy model’s estimation accuracy of prior probability distribution has become the core approach to improve compression performance. The closer the estimated distribution is to the actual distribution of $\tilde{y}_{i}$, the fewer bits are consumed during compression. As shown in Fig.~\ref{fig:general_structure}, our design combines local prior information $\Theta_i$ with global prior information $\theta_{j \le i}$ to generate the parameter for the estimated distribution. 
The local prior information is extracted from the $i$th viewport's quantized latent representation $\tilde{y}_i$ by any existing 2D image entropy model. The global prior information $\theta_{j \le i}$ is derived from multiple viewports' quantized latent representations, $[\tilde{y}_{0}, \tilde{y}_{1}, …, \tilde{y}_{j}]$, utilizing the proposed VPCT. 

\subsubsection{Motivation for VPCT}
\label{sec:motivation_for_vpct}
We now discuss the advancement of canonical entropy models shown in Fig.~\ref{fig:factorized} to \ref{fig:reference} to motivate the design of our VPCT-based entropy model. The prior probability distribution of existing entropy models are
\begin{equation*}
\begin{array}{ll}
\\
\text{Factorized~\cite{factorized}:} & \multicolumn{1}{c}{p_{\tilde{y}}(\tilde{y} \mid \psi)} \\
\text{} & \multicolumn{1}{c}{\Downarrow} \\
\text{Hyperprior~\cite{hyperprior}:} & \multicolumn{1}{c}{p_{\tilde{y}}\left(\tilde{y} \mid \theta_{h}\right)} \\
\text{} & \multicolumn{1}{c}{\Downarrow} \\
\text{Joint~\cite{joint}:} & \multicolumn{1}{c}{p_{\tilde{y}}\left(\tilde{y} \mid \theta_{h}, \theta_{c}\right)} \\
\text{} & \multicolumn{1}{c}{\Downarrow} \\
\text{Reference~\cite{reference}:} & \multicolumn{1}{c}{p_{\tilde{y}}\left(\tilde{y} \mid \theta_{h}, \theta_{c}, \theta_{r}\right)} \\\\
\end{array}
\end{equation*}
where factorized entropy model is an early work that takes a trainable parametric vector $\psi$ as prior information to approximate the probability of $\tilde{y}$, denoted ${p_{\tilde{y}}(\tilde{y} \mid \psi)}$. As shown in Fig.~\ref{fig:factorized}, since factorized entropy model dose not rely on any external input to estimate the probability distribution of $\tilde{y}_i$ and have a very simple structure, we omit the structure of factorized entropy model in the figure as in other studies~\cite{hyperprior, joint, reference}.
The hyperprior entropy model introduces hyperprior information $\theta_h$ by expanding the factorized entropy model with an additional branch, where $\tilde{z}_i$ is the quantized hyperprior latent obtained from $y_i$ by this branch. The hyperprior entropy model predicts the probability of $\tilde{y}_i$ according to $\theta_h$, denoted as $p_{\tilde{y}}\left(\tilde{y} \mid \theta_{h}\right)$.
Different from previous factorized entropy model, the hyperprior entropy model is adaptive to different image content with the help of the new hyperprior branch. 
As shown in Fig.~\ref{fig:hyperprior}, $\tilde{y}$ is modeled as zero-mean Gaussian distribution with predicted deviation $\tilde{\sigma}_i$, denoted as $p(\tilde{y}_i) \sim N \left(0, \sigma_i^{2}\right)$, where $\tilde{\sigma}_i$ is predicted by $\theta_h$. 
Then, the joint entropy model additionally includes image contextual information $\theta_c$ through an autoregressive context model (denoted as $CM$ in Fig.~\ref{fig:joint}), due to the property that the surrounding pixels of a given pixel have similar colors. In the context model, the quantized latent features within the convolution kernel are used to extract contextual information, thereby aiding the compression of remaining pixels. As shown in Fig.~\ref{fig:joint}, $\tilde{y}_i$ is modeled as Gaussian distribution with predicted mean $\tilde{\mu}_i$ and deviation $\tilde{\sigma}_i$, denoted as $p(\tilde{y}_i) \sim N \left(\tilde{\mu}_i, \tilde{\sigma}_i^{2}\right)$, where $\tilde{\mu}_i$ and $\tilde{\sigma}_i$ are predicted by jointing $\theta_h$ and $\theta_c$. The process of calculating the prior probability of $\tilde{y}_i$, by jointing parameters $\theta_h$ and $\theta_c$, can be denoted by $p_{\tilde{y}}(\tilde{y} \mid \theta_{h}, \theta_{c})$. 
Finally, the reference entropy model achieves SOTA performance by extending beyond the contextual information limited by kernel sizes. It adds the reference information $\theta_r$ extracted by the reference model (denoted as $R$ in Fig.~\ref{fig:reference}) to explore a wider range of quantized latent features to aid the processing of uncompressed pixels. 
After incorporating the $\theta_r$, the prior probability estimate for $\tilde{y}$ is given by $p_{\tilde{y}}\left(\tilde{y} \mid \theta_{h}, \theta_{c}, \theta_{r}\right)$.

While these four canonical entropy models mentioned above are widely used, they all are designed to process information from a single image and unable to capture the dependencies among different viewports. This leads to their inability to address the viewport discontinuity and overlapping challenges that increase bit consumption.
Therefore, we propose the VPCT module, which introduces global prior information of viewport correlation directly into the entropy model to enhance its prior probability distribution estimation and the codec's overall compression performance. Meanwhile, the VPCT module is designed to seamlessly integrate with existing entropy models, allowing it to capitalize on past developments and adapt to future advancements in 2D image compression structures.

\subsubsection{VPCT}
The key improvement of the proposed entropy model over existing entropy models is expressed as,
\begin{center}
$
p_{\tilde{y}_{i}}\left(\tilde{y}_{i} \mid \Theta_i \right) \Rightarrow p_{\tilde{y}_{i}}\left(\tilde{y}_{i} \mid \Theta_i,\theta_{j \le i}  \right)
$
\end{center}
where $p_{\tilde{y}_{i}}\left(\tilde{y}_{i} \mid \Theta_i \right)$ is the unified definition of canonical entropy models in Fig.~\ref{fig:entropy_models}, and $ p_{\tilde{y}_{i}}\left(\tilde{y}_{i} \mid \Theta_i,\theta_{j \le i}  \right)$ denotes the proposed entropy model. Unlike existing entropy models that only rely on local information $\Theta_i$ of an image, the proposed model introduces cross-image information $\theta_{j \le i}$ into the entropy model for the first time and can be detailed as,
\begin{center}
$
\begin{aligned}
p_{\tilde{y}_i}\left(\tilde{y}_i \mid \Theta_i, \theta_{j \le i} \right)&=\prod_{n}\left(\mathcal{N}\left(\tilde{\mu}_{i,n}, \tilde{\sigma}_{i,n}^{2}\right) * \mathcal{U}\left(-\frac{1}{2}, \frac{1}{2}\right)\right)\left(\tilde{y}_i\right)
\\
\text { with } \tilde{\mu}_{i,n}, \tilde{\sigma}_{i,n}&=f\left( \Theta_{i,n}, \theta_{j \le i}\right)
\text { and } \theta_{j \le i} = v\left(\tilde{y}_{j \le i}\right)
\end{aligned}
$
\end{center}\vspace{1em}
where $\tilde{y}_i$ is modeled as Gaussian with predicted mean $\tilde{\mu}_{i,n}$ and standard deviation $\tilde{\sigma}_{i,n}$, $f(\cdot)$ is the Fusion module in Fig.~\ref{fig:general_structure} that fuses local and global prior information to estimate the parameters of the Gaussian, and $v$ denotes the VPCT module. The local prior information $\Theta_i$ comes from the 2D codec integrated with VPCT. The global prior information $\theta_{j \le i}$ is extracted from the viewport latent representations $\tilde{y}_{j \le i}$ that have already been quantized. 

\begin{figure}
  \centering
  \includegraphics[width=0.8\linewidth]{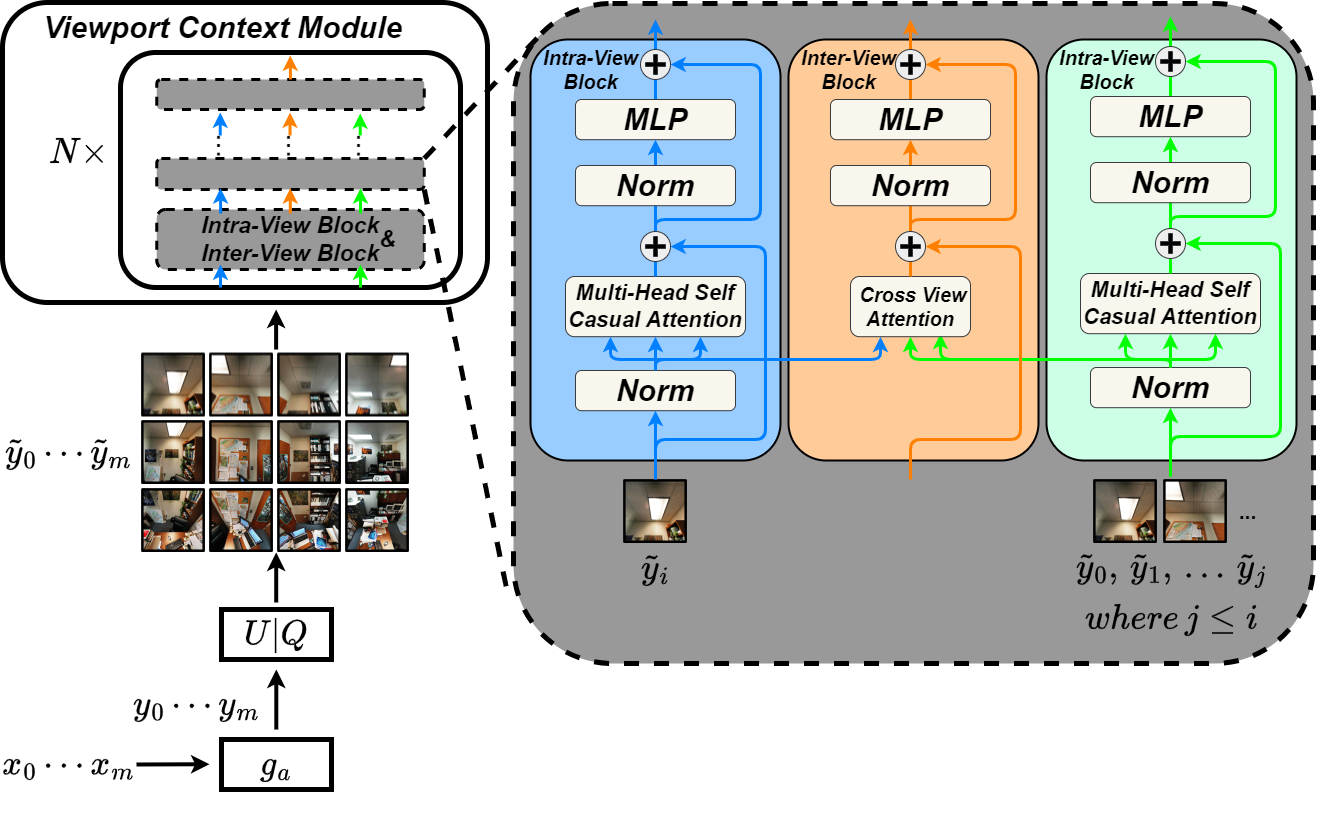}
  \caption{The overall structure of the VPCT module.}
  \label{fig:vpct}
\end{figure}

Fig.~\ref{fig:vpct} shows the structure of VPCT. For compatibility with 2D compression models, viewports are converted to latent representations individually via $g_a(\cdot)$ without sharing information with each other. 
In order to efficiently share information between viewports, the quantized latent representations of available viewports extracted from a 360\degree~image will be simultaneously fed into the VPCT. The VPCT contains $\textrm{N}$ basic transformer layers, each of which consists of an intra-view block and an inter-view block. 


\noindent
\textbf{Intra-view Block.} 
While causal contextual information in images, e.g., predicting a pixel based on many known pixels, has been shown effective for extracting the correlation within a quantized latent representation $\tilde{y}_i$, it cannot be captured by previous entropy models that have a limited range or only explore one previous pixel~\cite{joint, reference}. 
To capture the causal contextual information on a single viewport and thereby enhance the global correlation extraction across multiple viewports, we propose the transformer-based intra-view block.

As shown in Fig.~\ref{fig:vpct}, the intra-view block uses the structure of the basic layer of Vision Transformer (ViT)~\cite{vit}, consisting of layer normalization (LN), multiheaded self-attention layer (MSA), multi-layer perceptron (MLP), and residual connections. 
Unlike the self-attention design used in ViT, which calculates the correlation between every two positions within an image, we need to constrain the attention calculation to capture the causal contextual information within $\tilde{y}_i$. Specifically, we focus on calculating the correlation between the current position and the positions that have been quantized in the latent representation. We accordingly add an attention mask to the scaled dot product of queries and keys, i.e.,
\begin{center}
$
\operatorname{Attn}_{\operatorname{self}}(\boldsymbol{Q}, \boldsymbol{K}, \boldsymbol{V})=\operatorname{softmax}\left(\frac{\boldsymbol{Q K}^{T}}{\sqrt{d_{k}}}+\mathcal{M} \right) \boldsymbol{V}
$ \\
$
\mathcal{M}(w, h)=\left\{\begin{array}{ll}
0 & \text { if } w\le h \\
-\infty & \text { otherwise }
\end{array}\right. 
$
\end{center}\vspace{1em}
where queries $\boldsymbol{Q} \in \mathbb{R}^{N \times d_{k}}$, keys $\boldsymbol{K} \in \mathbb{R}^{N \times d_{k}}$ and values $\boldsymbol{V} \in \mathbb{R}^{N \times d_{v}}$ are projected from $S \in \mathbb{R}^{N \times C}$ which is a sequence of vectors reshaped from input viewport's quantized latent representation $\tilde{y}_{i} \in  \mathbb{R}^{H \times W \times C}$. 
The attention mask is denoted by $\mathcal{M} \in \mathbb{R}^{N \times N}$ where the value of each position is determined by its coordinates $(w,h)$. The construction process of $\mathcal{M}$ is summarized in Algorithm~\ref{alg:masks}(a).


\noindent
\textbf{Inter-view Block.} The inter-view block is designed to capture the correlation between current viewport's quantized latent $\tilde{y}_{i}$ with previous viewport's quantized latents $\tilde{y}_{j \le i}$. This design can help the model identify and remove the overlapping area between different viewports. We design a new attention layer to support the inter-view block. Specifically, the queries set $\boldsymbol{Q}_i \in \mathbb{R}^{N \times d_{k}}$ is projected from a sequence of vectors reshaped from $\tilde{y}_{i}$, whereas the keys set 
$\boldsymbol{K}_{j \le i} \in  \mathbb{R}^{\left ( i+1 \right ) \cdot N \times d_{k}}$
and the values set $\boldsymbol{V}_{j \le i} \in  \mathbb{R}^{\left ( i+1 \right ) \cdot N \times d_{v}}$ are projected from multiple reshaped viewport latent $\tilde{y}_{j \le i}$. Therefore, the cross-view attention used in the inter-view block can be formulated as follows,
\begin{center}\vspace{1em}
$
\operatorname{Attn}_{\operatorname{cross}}(\boldsymbol{Q_i}, \boldsymbol{K_{j \le i}}, \boldsymbol{V_{j \le i}})=\operatorname{softmax}\left(\frac{\boldsymbol{Q_{i} K^{T}_{j \le i}}}{\sqrt{d_{k}}}+\mathcal{M}_{cross} \right) \boldsymbol{V_{j \le i}}
$
\end{center}\vspace{1em}
where $\mathcal{M}_{cross} \in \mathbb{R}^{N \times \left ( i+1 \right ) \cdot N}$ denotes the attention mask used in the cross-view attention layer. Since the keys and queries have different shapes in the cross-view attention, the shape of $\mathcal{M}_{cross}$ differs from that of $\mathcal{M}$. 
The construction process of $\mathcal{M}_{cross}$ is shown in Algorithm~\ref{alg:masks}(b), and the value at each position in $\mathcal{M}_{cross}$ is
\begin{center}
$
\mathcal{M}_{cross}(w, h)=\left\{\begin{array}{ll}
0 & \text { if } w-i \cdot N \le h \\
-\infty & \text { otherwise }
\end{array}\right.
$
\end{center}\vspace{1em}
The quantized latent representations of viewports are fed to the inter-view block sequentially. Each viewport can take the earlier viewports as prior information to enhance the estimation accuracy of the entropy model, making the compression of the later viewports more effective.

\begin{algorithm}[H]
\caption{Construction of attention masks}
\label{alg:masks}
\centering

\begin{minipage}[t]{0.48\linewidth}
\vspace{0pt}
\textbf{(a) Intra-view mask $\mathcal{M}(w,h)$}
\begin{algorithmic}[1]
\Require $N$
 \Ensure Zero matrix $\mathcal{M}\in\mathbb{R}^{N\times N}$
\For{$w\gets 0$ \textbf{to} $N-1$}
    \For{$h\gets 0$ \textbf{to} $N-1$}
        \If{$h < w$}
            \State $\mathcal{M}(w,\;h)\gets -\infty$
        \EndIf
    \EndFor
\EndFor 
\State Return $\mathcal{M}$

\end{algorithmic}
\end{minipage}
\hfill
\begin{minipage}[t]{0.48\linewidth}
\vspace{0pt}
\textbf{(b) Inter-view mask $\mathcal{M}_{\text{cross}}(w,h)$}
\begin{algorithmic}[1]
\Require $N$ and $i$
\Ensure Zero matrix $\mathcal{M}_{\text{cross}}\in\mathbb{R}^{N\times (i+1)\cdot N}$
\For{$w \gets 0$ \textbf{to} $(i+1)\cdot N-1$}
    \For{$h \gets 0$ \textbf{to} $N-1$}
        \If{$h < w - i \cdot N$}
            \State $\mathcal{M}_{cross}(w,\;h)\gets -\infty$
        \EndIf
    \EndFor    
\EndFor
\State Return $\mathcal{M}_{cross}$
\end{algorithmic}
\end{minipage}
\end{algorithm}


\noindent\textbf{Feature Fusion Layer.}
To integrate the proposed VPCT with existing 2D image compression structures, as depicted in Fig.~\ref{fig:general_structure}, the feature fusion layer is employed to combine the global prior information $\theta_{j \le i}$ generated by VPCT with local prior information $\Theta_i$ to estimate the parameters for the entropy model. Specifically, $\Theta_{i} \in \mathbb{R}^{H \times W \times C_l}$ and $\theta_{j \le i} \in \mathbb{R}^{H \times W \times C_g}$ are concatenated along the channel dimension. Subsequently, this input feature undergoes three $1 \times 1$ convolutional layers with two leaky ReLU activation~\cite{leakyrelu} in between, resulting in the Gaussian parameters $\mu$ and $\sigma$. For example, to integrate the Reference model in Fig.~\ref{fig:reference}, the captured global prior information between viewports is concatenated with reference information, contextual information, and hyperprior information to estimate the Gaussian parameters via the feature fusion layer. The other compression models~\ref{fig:factorized}-\ref{fig:joint} will also be integrated with VPCT in the same way. 


\noindent\textbf{Loss Function.}
Similar to previous learning-based compression, the proposed viewport codec supervises model training by jointly minimizing the bitstream length (rate) and the distortion of the reconstructed viewport, i.e.,

\begin{center}\vspace{1em}
$
\begin{aligned}
\mathcal{L}= & \mathcal{R}(\tilde{\boldsymbol{y}}_i)+\mathcal{R}(\tilde{\boldsymbol{z}}_i)+\lambda \cdot \mathcal{D}(\boldsymbol{x}_i, \tilde{\boldsymbol{x}}_i) \\
= & \underbrace{\mathbb{E}\left[-\log _{2}\left(p_{\tilde{\boldsymbol{y}}_i}(\tilde{\boldsymbol{y}}_i \mid \Theta_i,\theta_{j \le i})\right)\right]+\mathbb{E}\left[-\log _{2}\left(p_{\tilde{\boldsymbol{z}}_i \mid \boldsymbol{\psi}}(\tilde{\boldsymbol{z}}_i \mid \boldsymbol{\psi})\right)\right]}_{\text{rate}} \\
& +\lambda \cdot \underbrace{\mathcal{D}(\boldsymbol{x}_i, \tilde{\boldsymbol{x}}_i)}_{\text {distortion}}
\end{aligned}
$
\end{center}\vspace{1em}
where $\lambda$ is a coefficient for balancing rate and distortion.

\section{Experiment}
\label{sec:exp}

\subsection{Experiment Setting}
\label{sec:exp:imp}
\subsubsection{Training Details}
All experiments are conducted using CompressAI~\cite{compressai}, an open-source library designed for developing and evaluating learning-based neural image codecs. We employ the largest public dataset for 360\degree~image compression~\cite{structure_map}, which contains 19,590 training images, to train our proposed model.
The proposed codec is trained with various $\lambda \in \left \{ 0.0018, 0.0035, 0.0075, 0.013, 0.025, 0.048 \right \}$ to generate rate-distortion curves. The loss function employs mean squared error (MSE) for calculating distortion. We utilize a batch size of 8 for training each model. Unless specified otherwise, all models are optimized using the Adam optimizer for 200 epochs with a learning rate of $10^{-4}$.

\subsubsection{Viewport Extraction} By default, we set horizontal and vertical FoV as 90\degree~to extract viewports from a single 360\degree~image. 
We set the viewport centers at \{(0.0\degree, 0.0\degree), (90.0\degree, 0.0\degree), (180.0\degree, 0.0\degree), (270.0\degree, 0.0\degree), (0.0\degree, 90.0\degree), (0.0\degree, -90.0\degree)\} to cover the entire sphere.
To evaluate the impact of viewport extraction settings, we report the results of 5 different settings in ablation studies, including FoV size of {45\degree, 67.5\degree, 90\degree, 112.5\degree, 135\degree}. We emphasize that viewports in 360\degree~image compression are referred to as a set of 2D windows that must all be offline compressed in one pass. The concept is different from 360\degree~video streaming, where a single viewport is extracted for streaming and viewing. Hence, settings of viewing trajectory, user viewing patterns, or head-mounted displays do not affect the compression process.

\subsubsection{Evaluation Datasets} To evaluate the compression efficiency of the proposed method, we test it on several datasets with different resolutions and visual content, including the LIC360 testing set (200 images with resolution of 512×1024)~\cite{structure_map}, Flickr360 (50 images with resolution of 1024×2048)~\cite{flickr360}, CVIQ (16 images with resolution of 2048×4096)~\cite{cviq}, and SaliencyVR, where we use three images (with resolution of 4096×8192) for testing~\cite{saliencyvr}.

\subsubsection{Evaluation Metrics}
We utilize the rate-distortion{-perception} metrics to measure the effectiveness of the proposed pipeline. As different viewport extraction settings introduce different total numbers of pixels on extracted viewports, utilizing bits per pixel to calculate the bit consumption is unfair. Thus, we employ bits per image (BPI) to measure the bits consumption of compressing the raw 360\degree~image. 
{To quantify the quality of the reconstructed images, we leverage the viewport-based peak signal-to-noise ratio (V-PSNR)~\cite{structure_map,vpsnr} to measure the distortion, and we also report the viewport-based structural similarity index (V-SSIM)~\cite{ssim} and the viewport-based learned perceptual image patch similarity (V-LPIPS)~\cite{lpips} to evaluate the perceptual quality of the reconstructed images.}

\subsection{Performance Comparison}
\label{sec:exp:com}

To demonstrate the effectiveness of the proposed viewport-based neural compression pipeline, we conduct a comprehensive analysis with three distinct categories of image compression methods -- a) the learning-based 360\degree~image compression with state-of-the-art performance, denoted by SOTA~\cite{structure_map}, b) learning-based 2D image codecs, including Factorized~\cite{factorized}, Hyperprior~\cite{hyperprior}, Joint~\cite{joint}, and Reference~\cite{reference}, and c) hand-crafted legacy codecs, VTM and WebP. Both learning-based 360\degree~and 2D codecs are trained and tested on ERP images, following the conventional pipeline. For hand-crafted codecs, we apply them on both the ERP and viewport formats, denoted by `erp` and `vp`, to ensure fair comparisons. The proposed codec using the viewport-based pipeline is constructed upon the Reference model.

\begin{figure*}[h]
\centering
\setlength{\tabcolsep}{2pt}
\renewcommand{\arraystretch}{1.0}
\begin{tabular}{c cccc}
    & LIC360 (1K)~\cite{structure_map} & Flickr360 (2K)~\cite{flickr360} & CVIQ (4K)~\cite{cviq} & SaliencyVR (8K)~\cite{saliencyvr} \\

    \raisebox{9.ex}{\rotatebox[origin=c]{90}{V-PSNR}} &
    \includegraphics[width=0.23\linewidth]{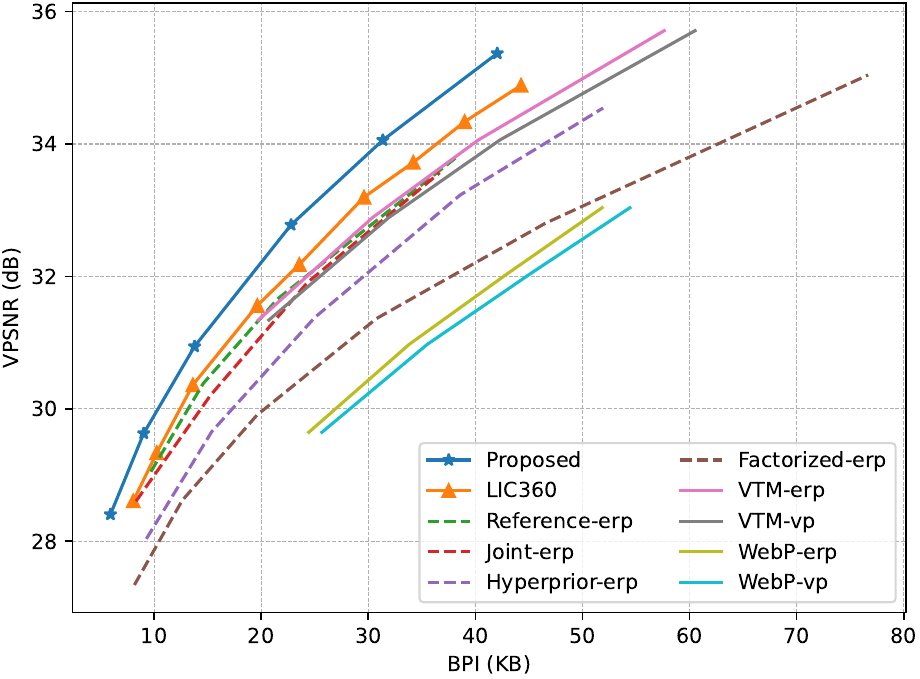} &
    \includegraphics[width=0.23\linewidth]{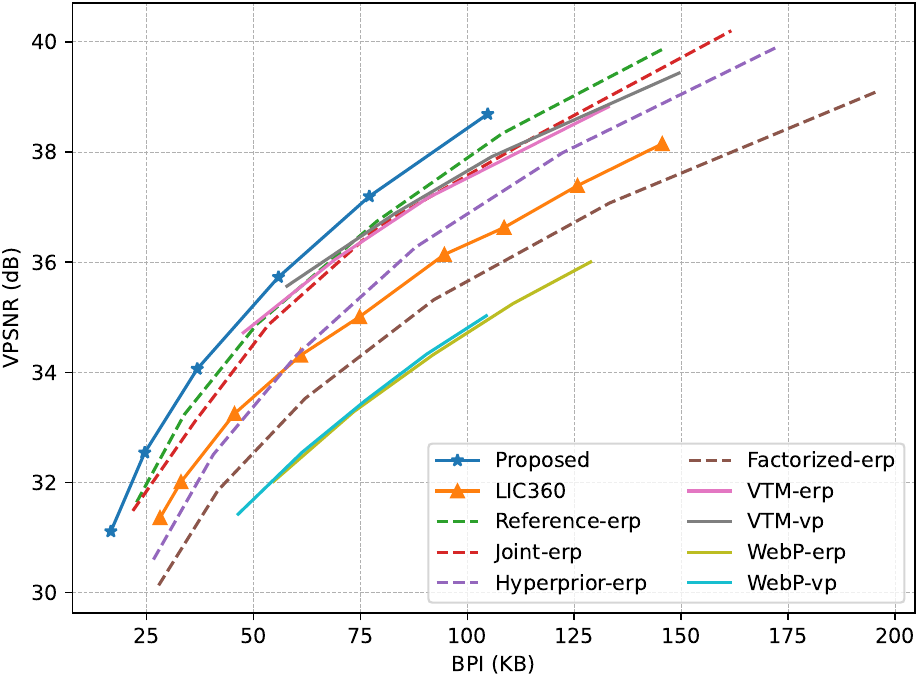} &
    \includegraphics[width=0.23\linewidth]{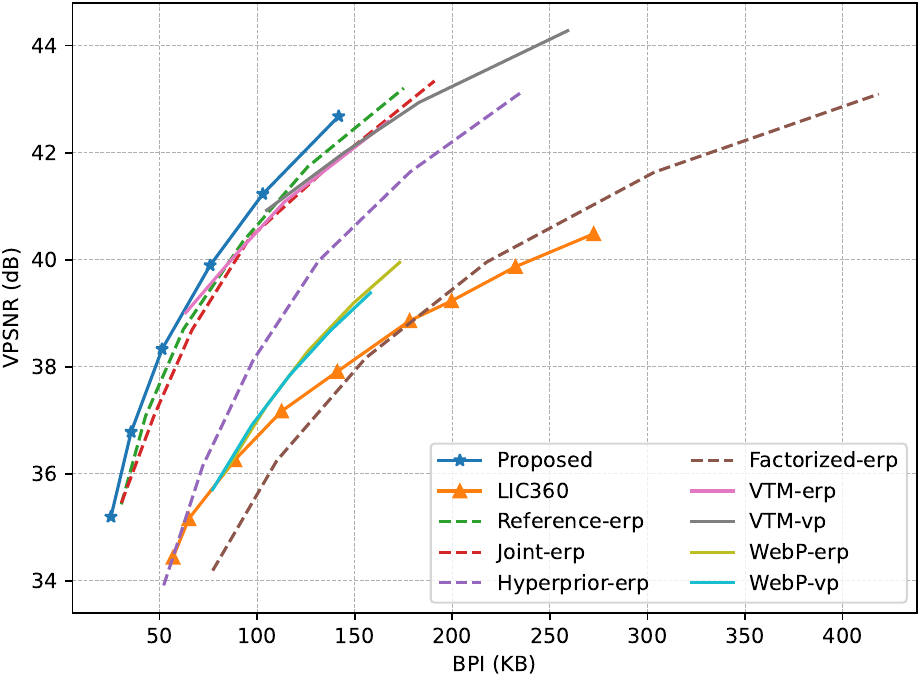} &
    \includegraphics[width=0.23\linewidth]{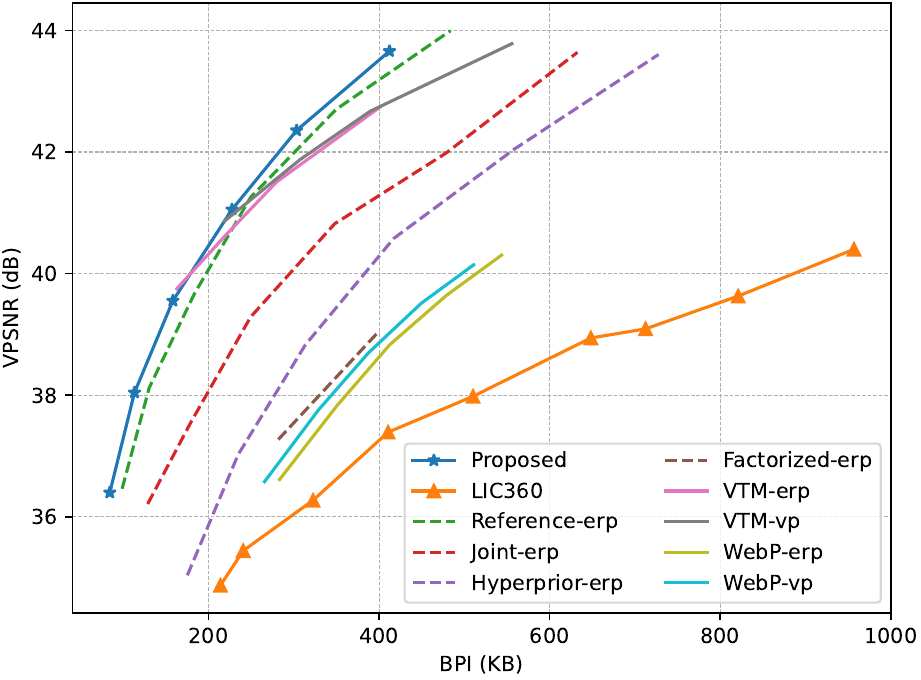} \\

    \raisebox{9.ex}{\rotatebox[origin=c]{90}{V-SSIM}} &
    \includegraphics[width=0.23\linewidth]{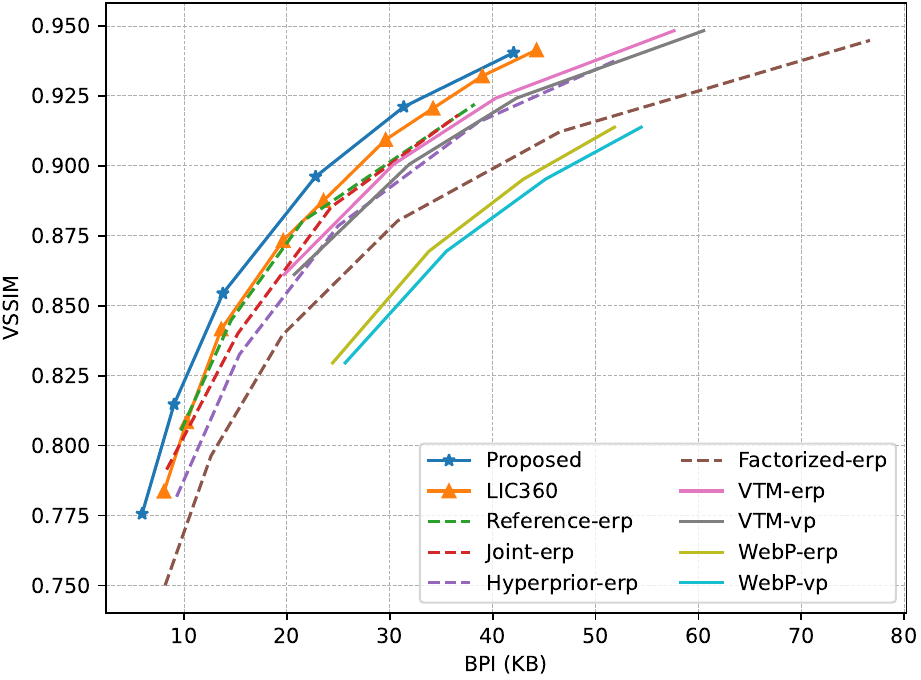} &
    \includegraphics[width=0.23\linewidth]{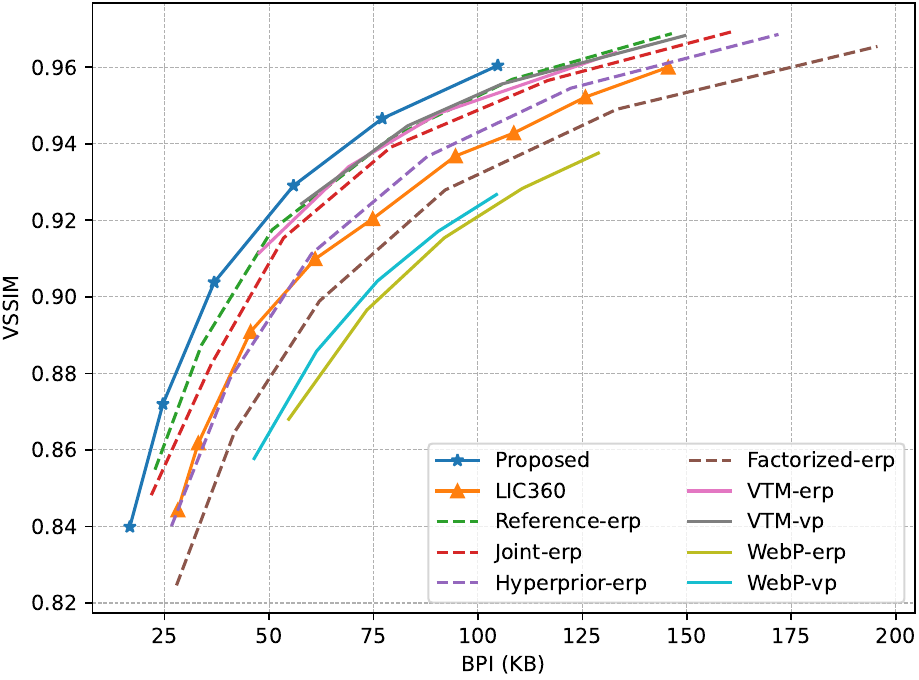} &
    \includegraphics[width=0.23\linewidth]{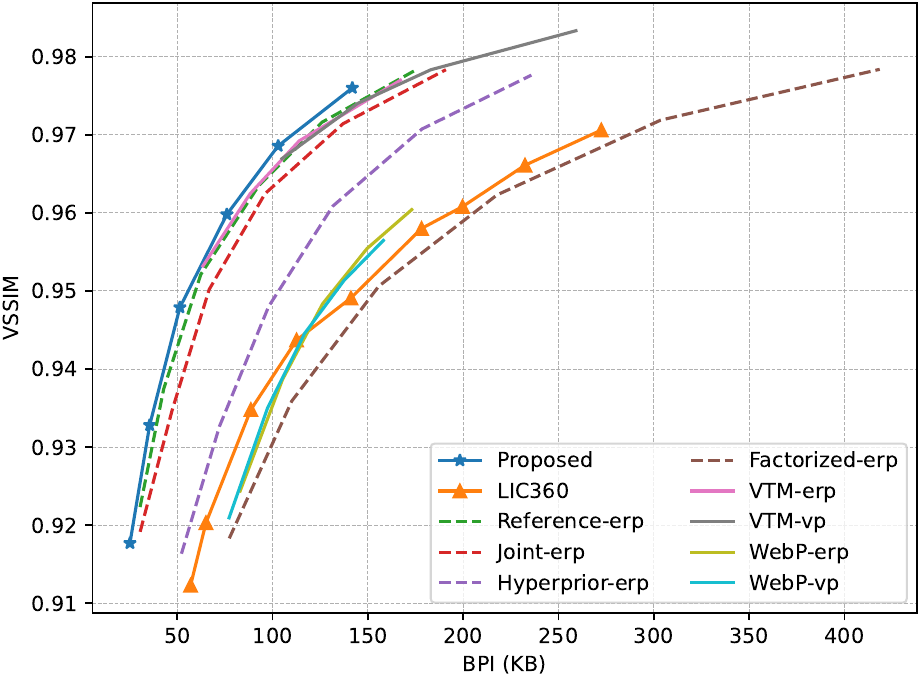} &
    \includegraphics[width=0.23\linewidth]{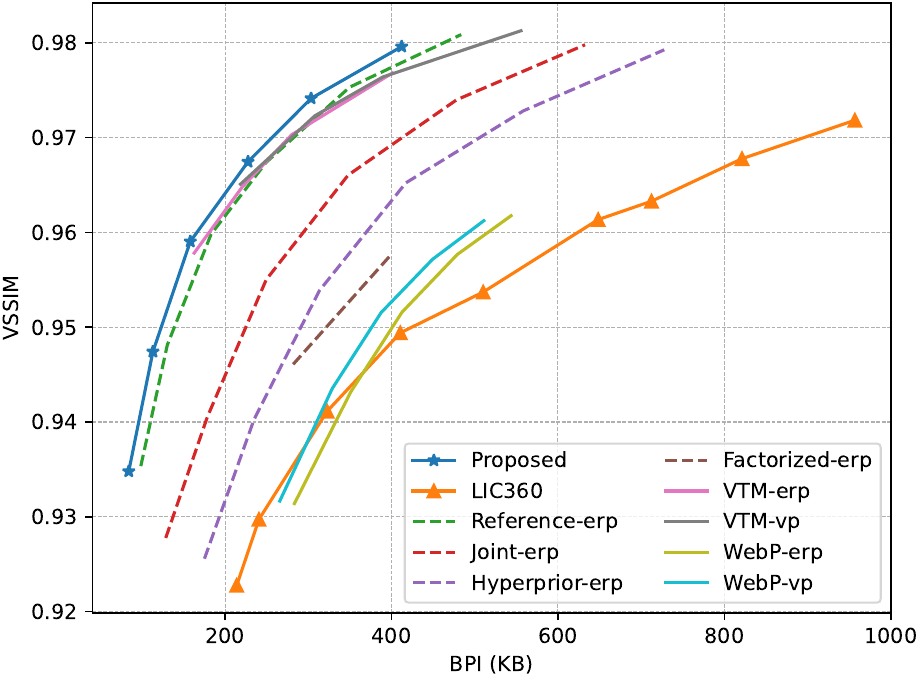} \\

    \raisebox{9.ex}{\rotatebox[origin=c]{90}{V-LPIPS}} &
    \includegraphics[width=0.23\linewidth]{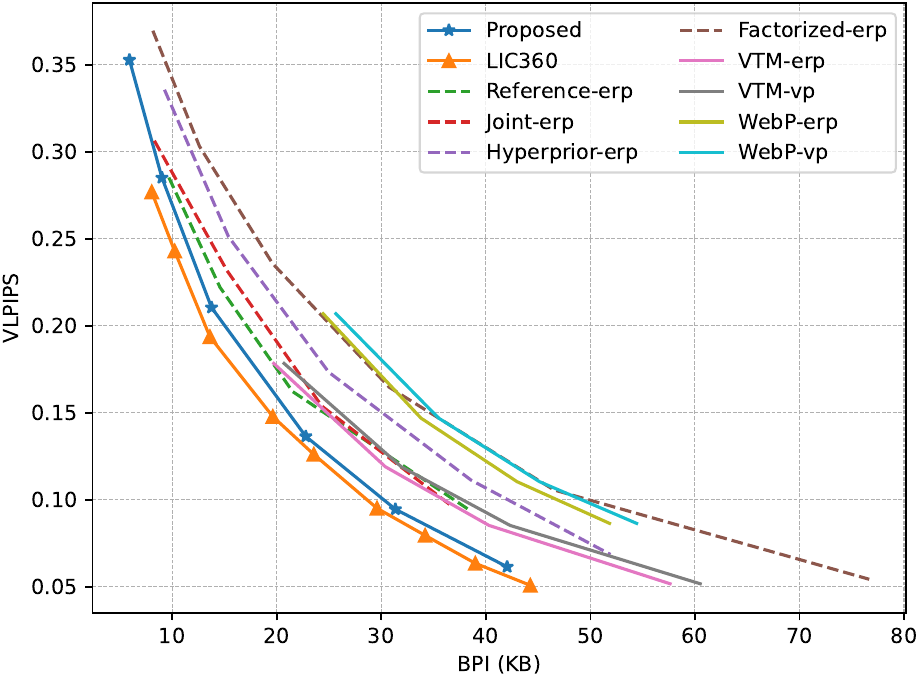} &
    \includegraphics[width=0.23\linewidth]{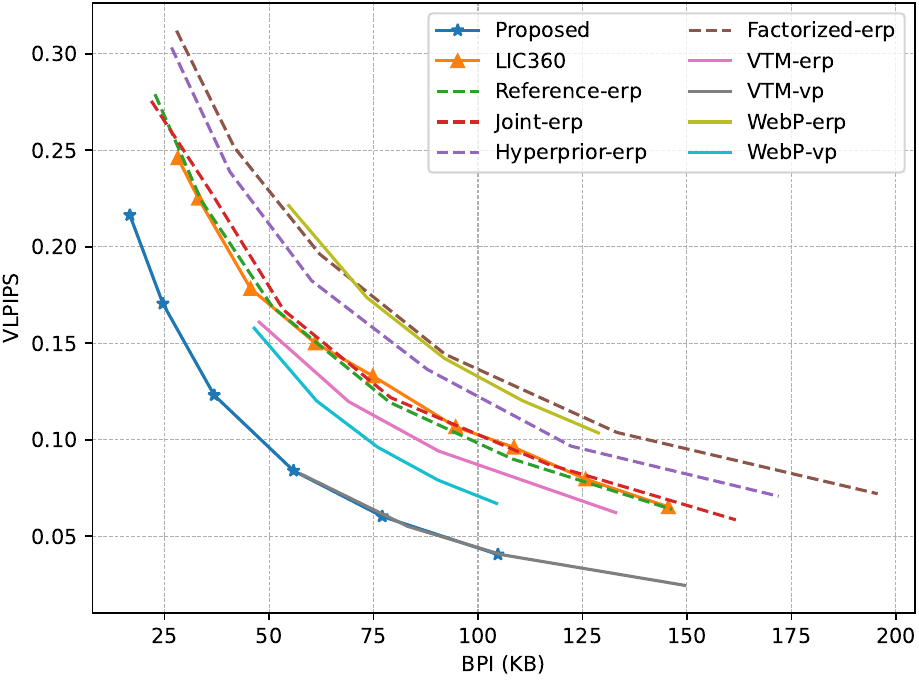} &
    \includegraphics[width=0.23\linewidth]{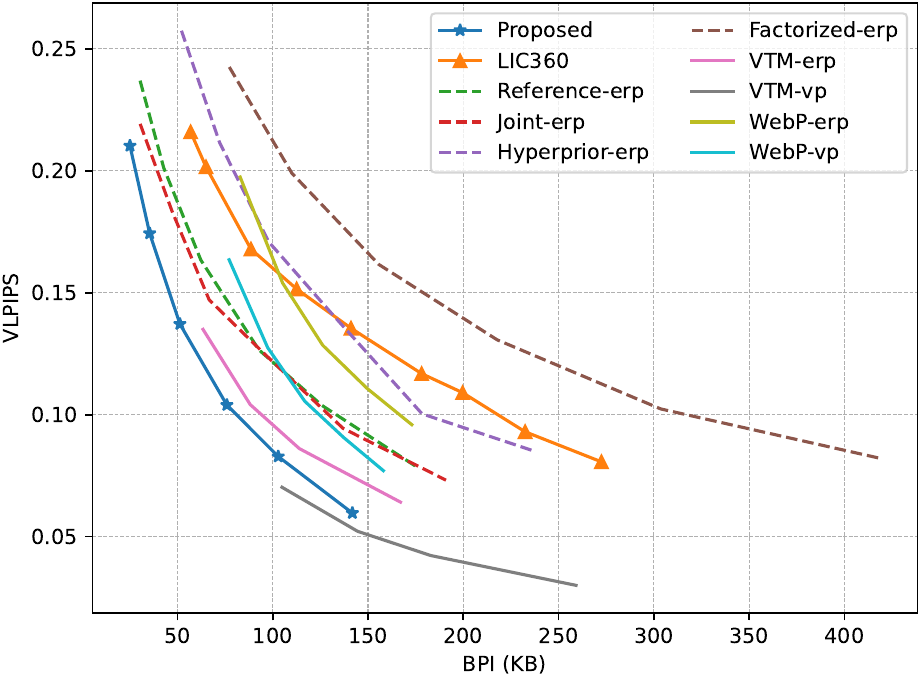} &
    \includegraphics[width=0.23\linewidth]{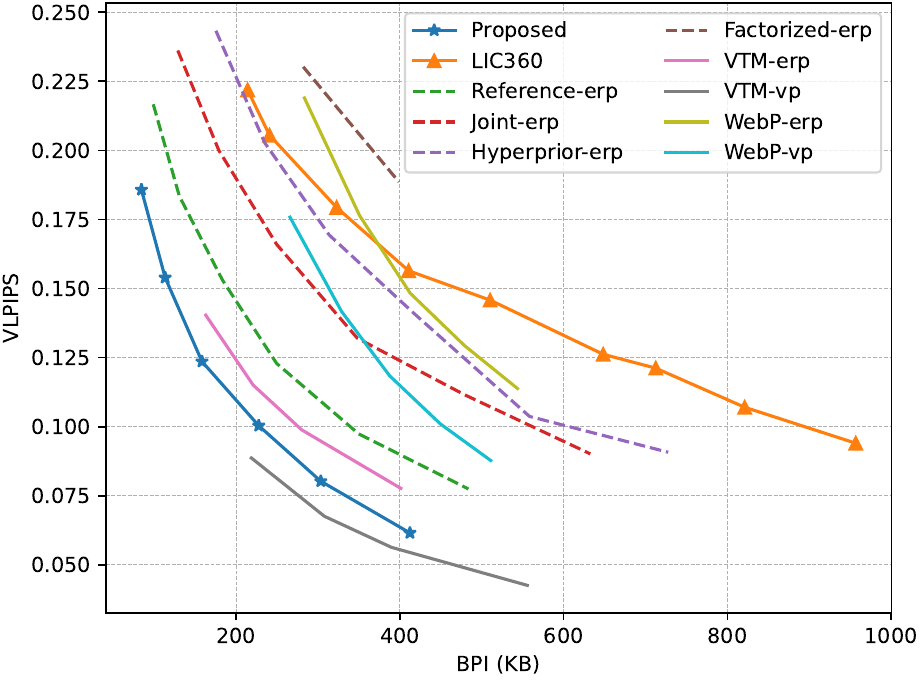}
\end{tabular}
\caption{
{Compression performance comparison.}
}
\label{fig:exp:overall-comparison}
\end{figure*}

{
Fig.~\ref{fig:exp:overall-comparison} shows the rate-distortion-perception performance of our method on four test datasets with different resolutions and visual characteristics. On the LIC360 testing set, the proposed viewport-based pipeline consistently outperforms the conventional ERP-based pipeline, achieving an average bitrate saving of $14.01\%$ compared to LIC360 codec under the same V-PSNR (see first row and first column of Fig.~\ref{fig:exp:overall-comparison}). Meanwhile, our method outperforms the baselines on the other three test datasets with higher resolutions. However, the conventional 360\degree~image SOTA codec cannot generalize well to these datasets, and its performance is even worse than the hand-crafted codecs on 8K SaliencyVR dataset (see last column of Fig.~\ref{fig:exp:overall-comparison}). In our viewport-based pipeline, the viewport extraction minimizes oversampling and geometric distortion during image projection, while the viewport codec maximally compresses the overlapping viewports through the VPCT, both of which contribute to the performance gains.}



\subsection{Integration with Existing 2D Codecs}
\label{sec:exp:ext}
We now integrate the proposed viewport-based neural 360\degree~image compression pipeline with four learning-based 2D image codecs to validate its effectiveness. We evaluate our pipeline with and without VPCT to dissect the performance gain brought by the viewport extraction and the viewport codec.
For benchmarking, we also evaluate the 2D codecs in the conventional pipeline using ERP image projection. 
\begin{figure}
    \centering
    \begin{minipage}[c]{\linewidth}
        \centering
        \subfigure[Factorized]{
            \includegraphics[width=0.472\linewidth]{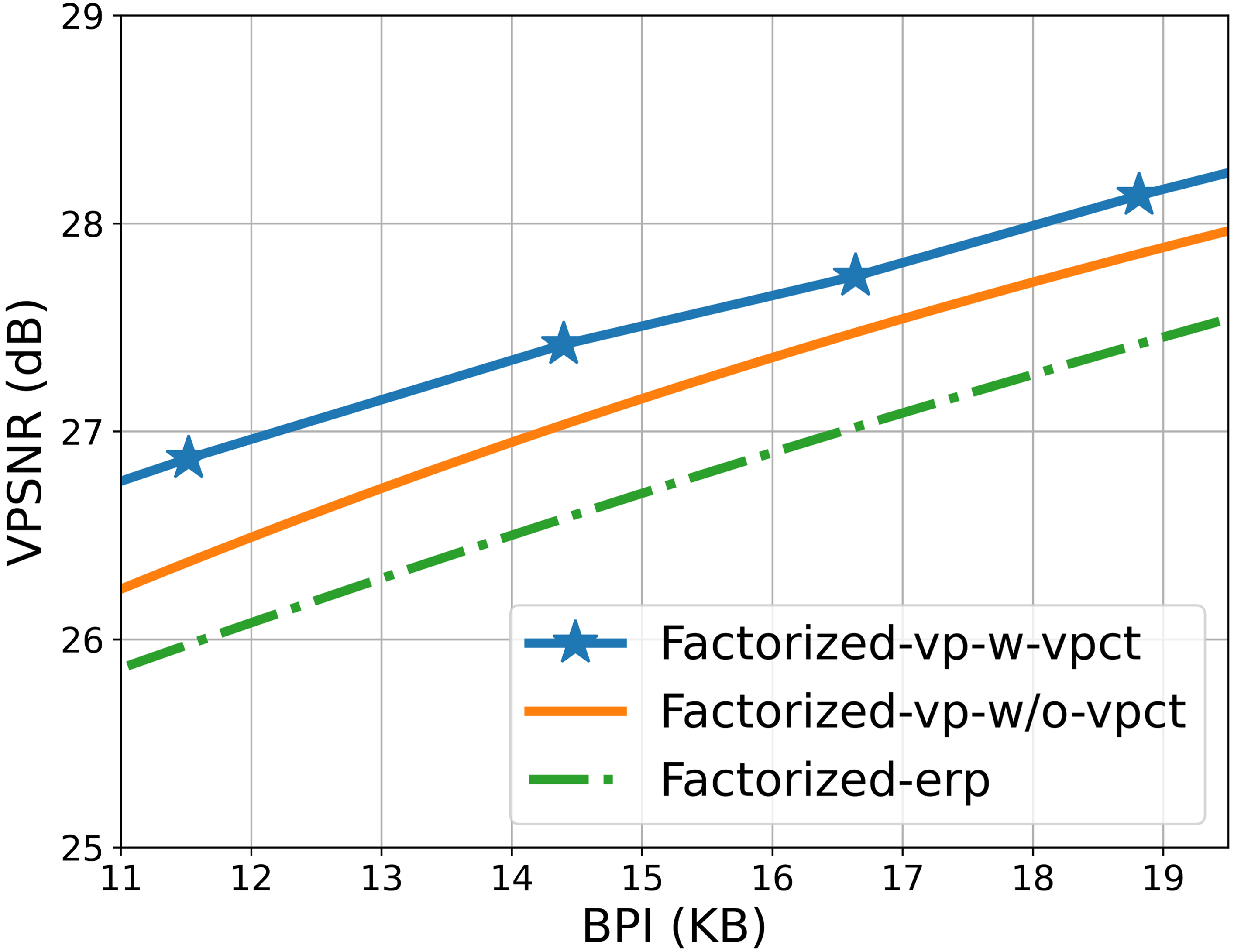}
            \label{fig:exp:factor}
        }
        \subfigure[Hyperprior]{
            \includegraphics[width=0.472\linewidth]{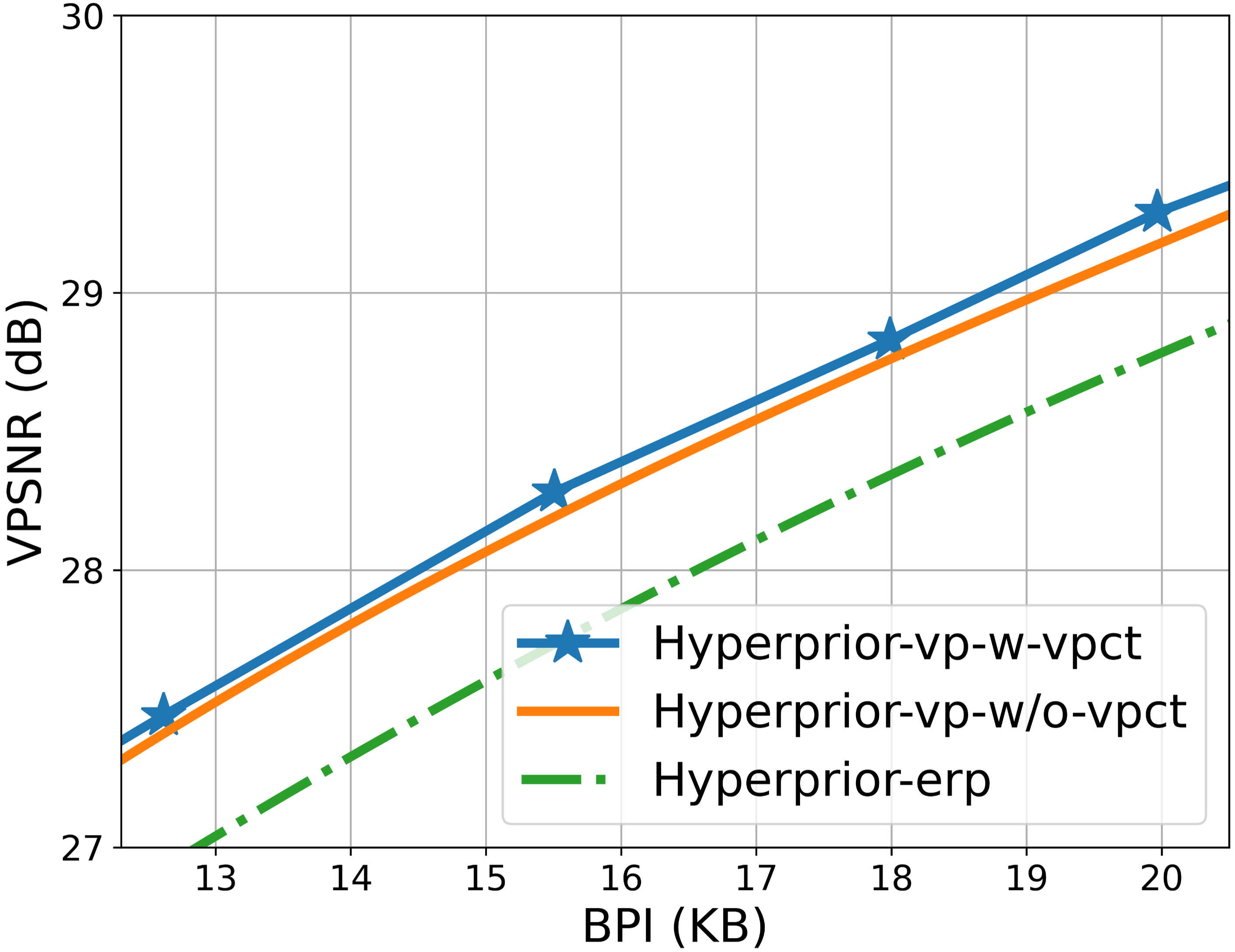}
            \label{fig:exp:hyper}
        }
    \end{minipage}\\
    \begin{minipage}[c]{\linewidth}
        \centering
        \subfigure[Joint]{
            \includegraphics[width=0.472\linewidth]{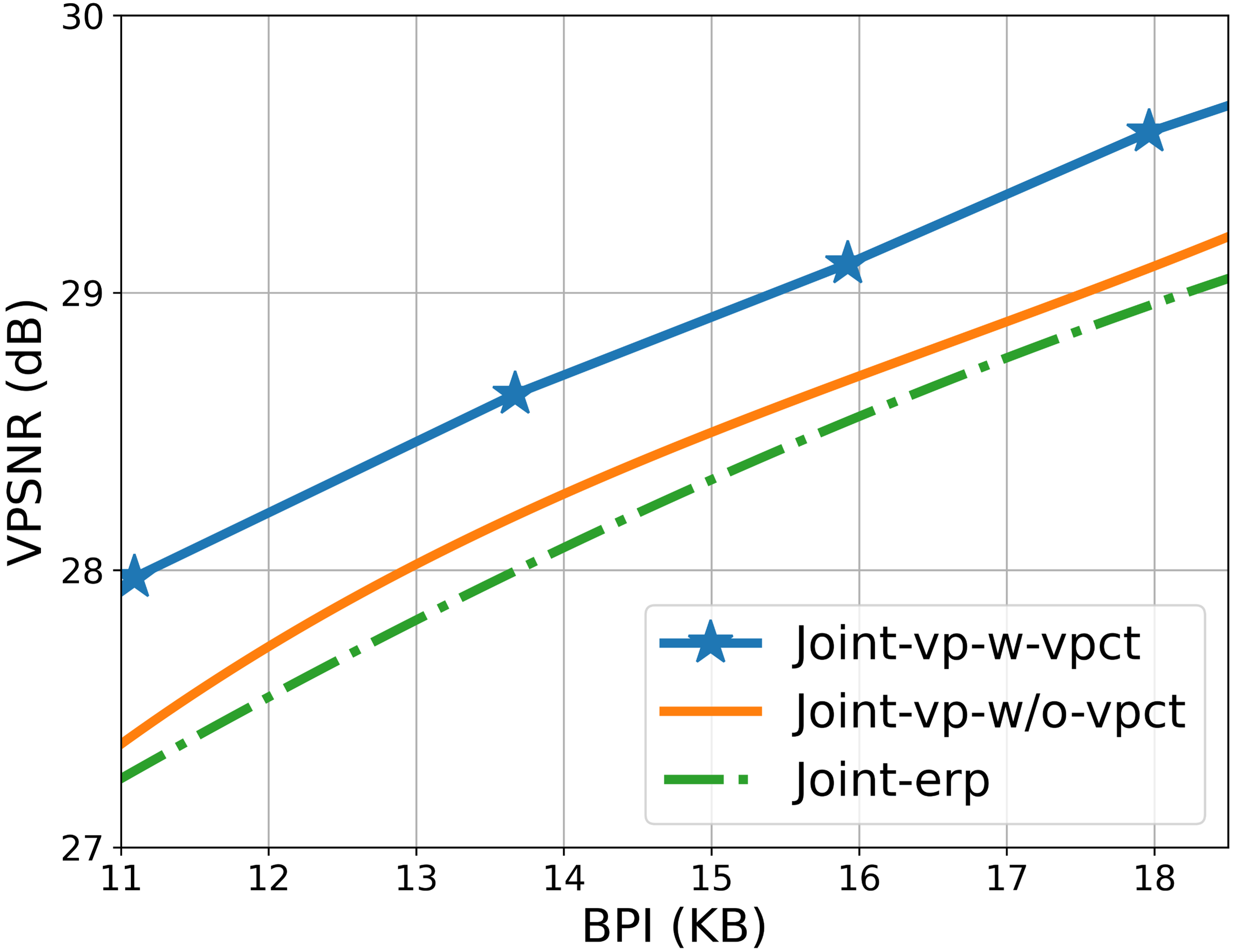}
            \label{fig:exp:joint}
        }
        \subfigure[Reference]{
            \includegraphics[width=0.472\linewidth]{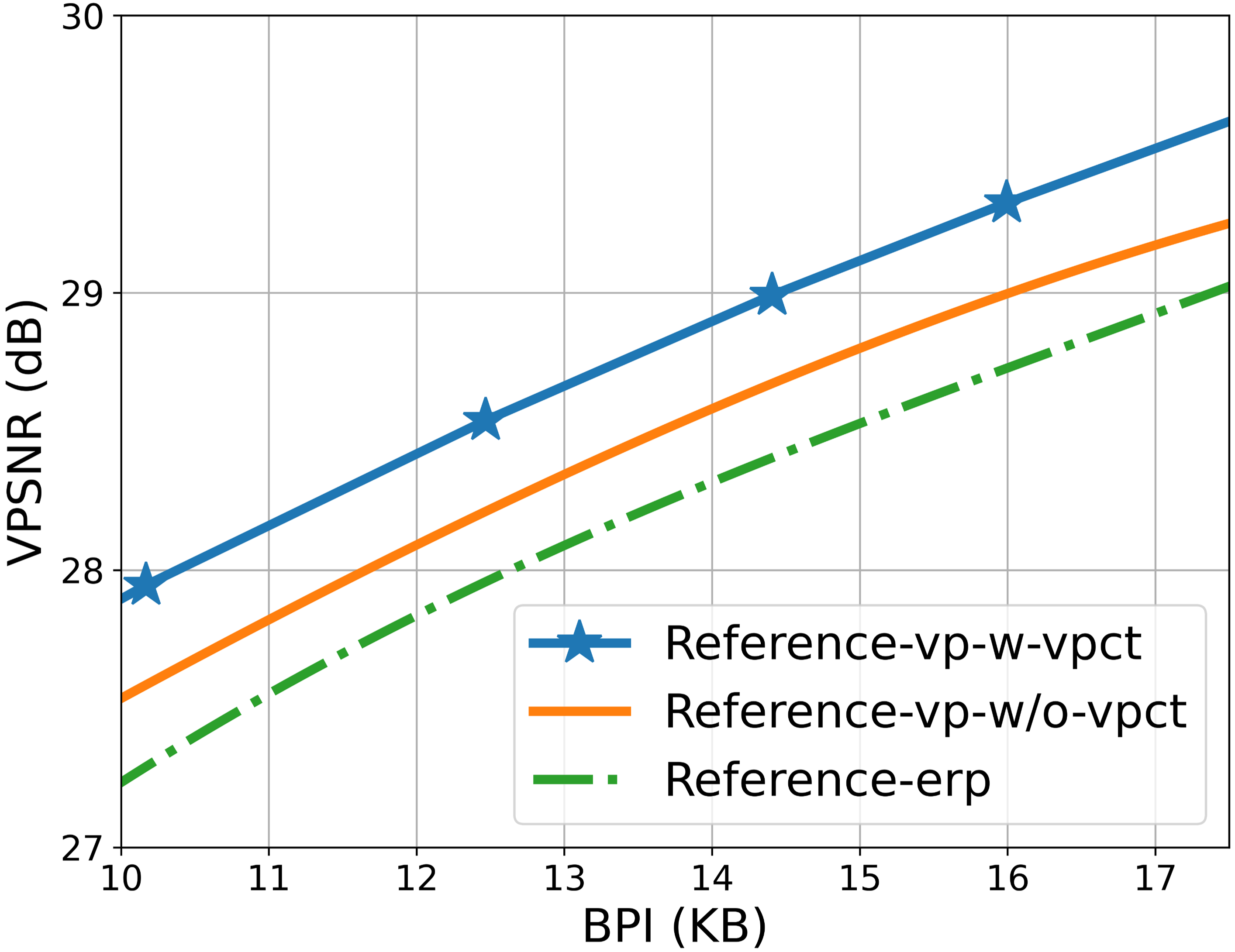}
            \label{fig:exp:ref}
        }
    \end{minipage}
    \caption{Compression performance comparison of integrating our viewport-based pipeline with existing 2D image codecs.}
    \label{fig:exp:r2f}
\end{figure}

Fig.~\ref{fig:exp:r2f} shows that the proposed pipeline achieves significant improvement over the conventional ERP-based pipeline across various underlying 2D codecs. Moreover, both the viewport extraction and the VPCT-based neural viewport codec bring noticeable gains. In Fig.~\ref{fig:exp:factor}, our pipeline outperforms the conventional one because it introduces viewport content information into the factorized entropy model which by default only relies on content-unrelated parameters for prior probability distribution estimation (see Secction~\ref{sec:motivation_for_vpct}).
Similarly, in Fig.~\ref{fig:exp:hyper}-~\ref{fig:exp:ref}, our pipeline exhibits performance gains on top of the other models. 

\begin{figure*}[htbp]
  \centering
  \centering
    \subfigure[Hyperprior]{
        \includegraphics[width=0.3\linewidth]{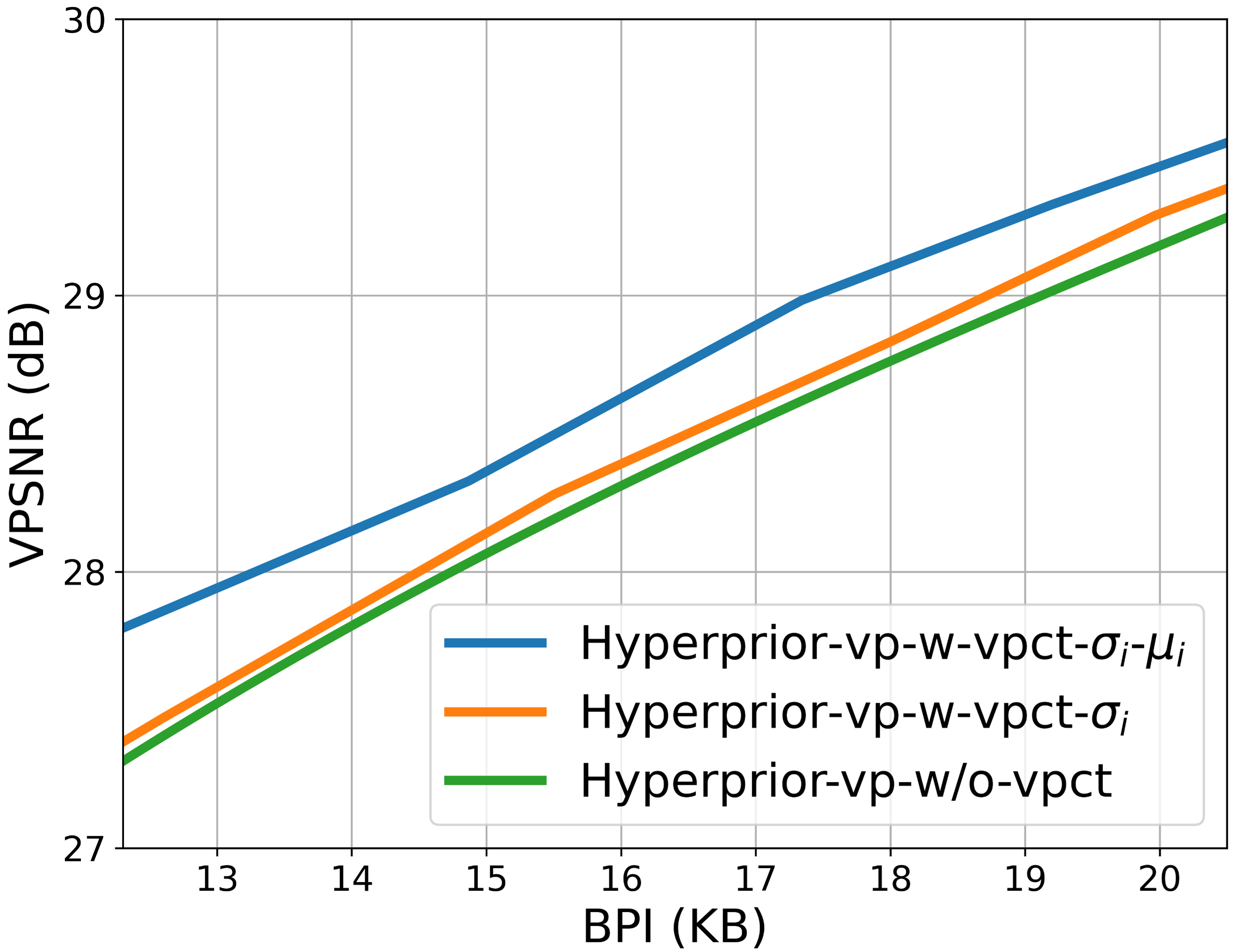}
        \label{fig:hyperprior-gaussian}
    }
  \centering
    \subfigure[Joint]{
        \includegraphics[width=0.3\linewidth]{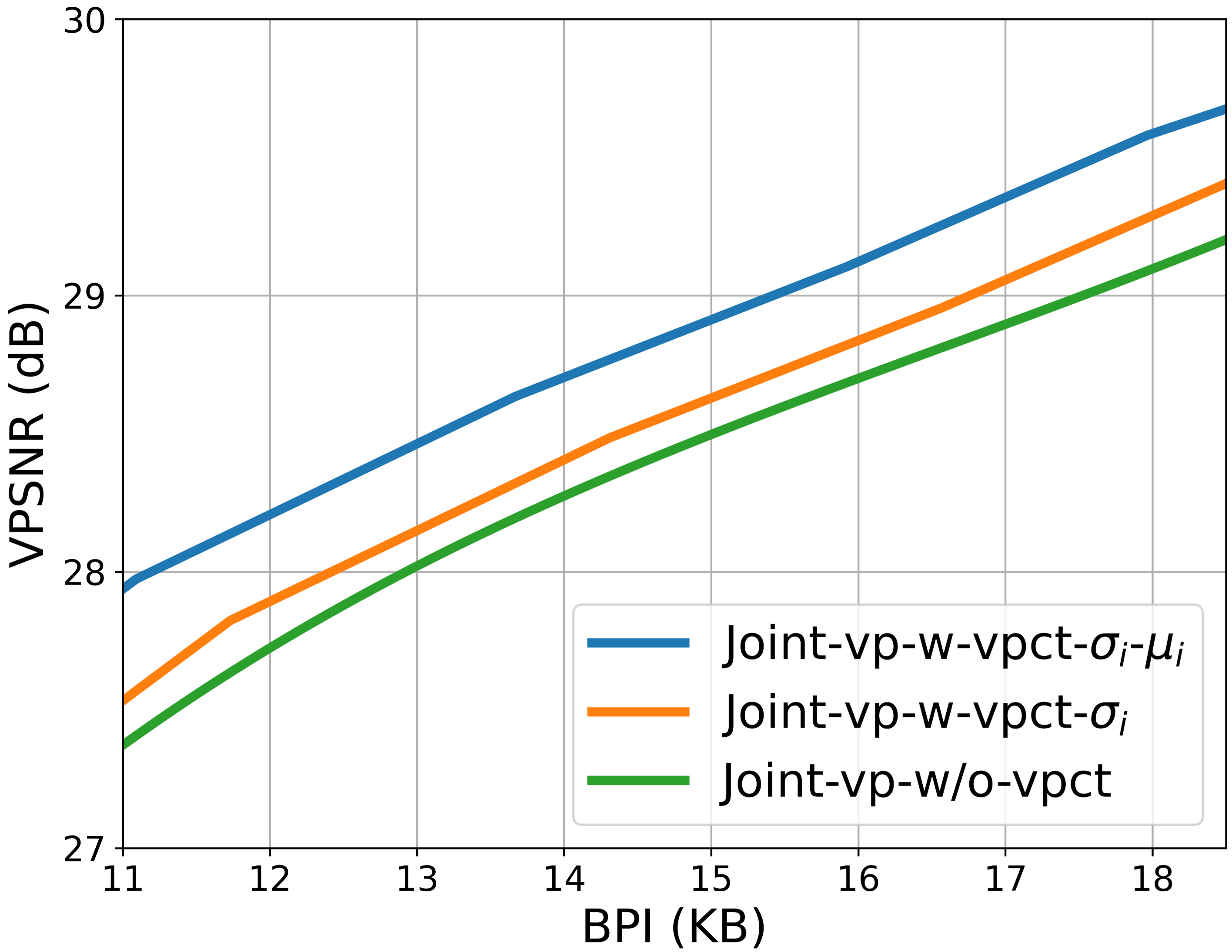}
        \label{fig:joint-gaussian}
    }
    \subfigure[Reference]{
        \includegraphics[width=0.3\linewidth]{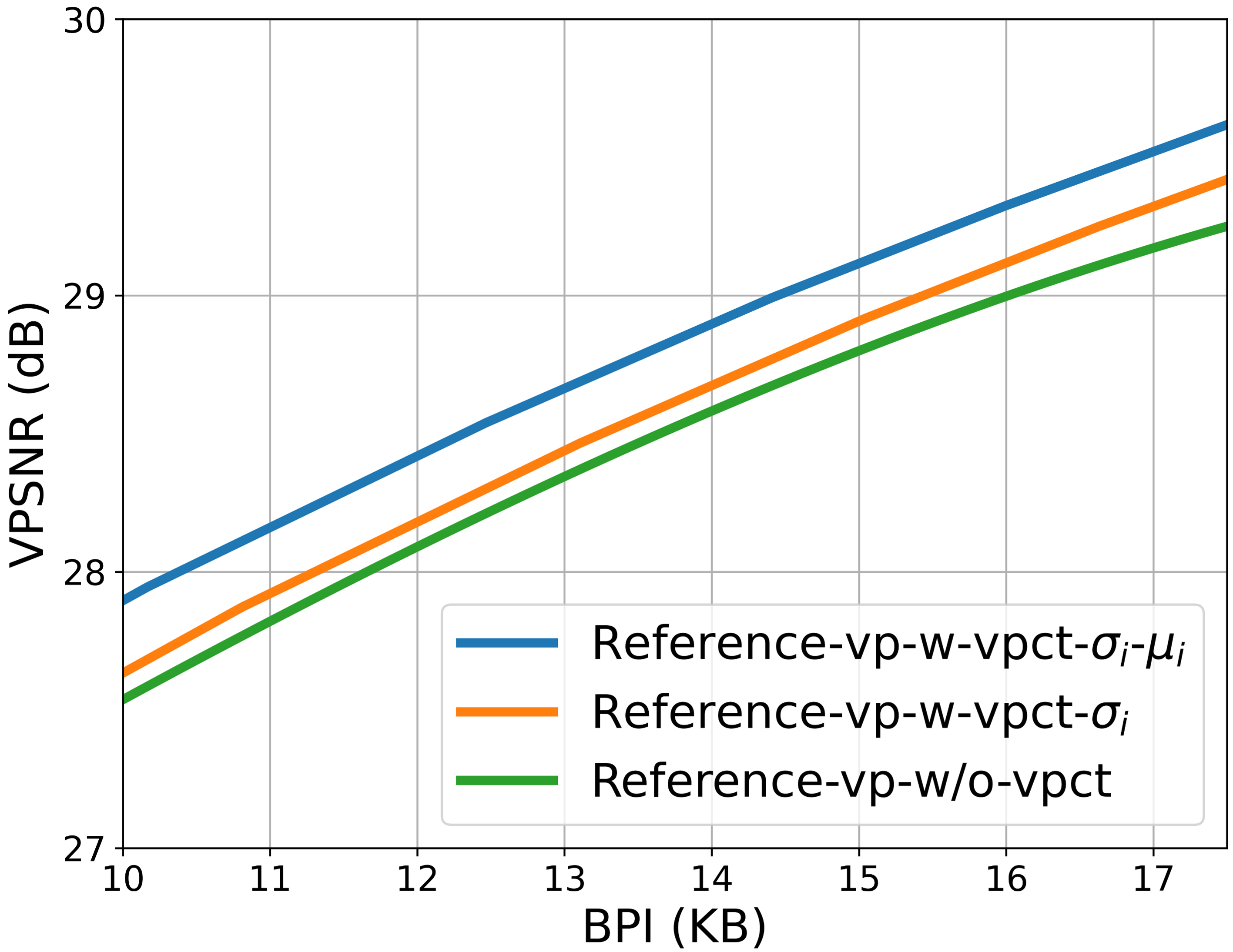}
        \label{fig:reference-gaussian}
    }
  \caption{
  {Analysis of gaussian parameters}}
  \label{fig:exp:ana_gaussion_params}
\end{figure*}

We also observe in Fig.~\ref{fig:exp:r2f} that the Hyperprior model exhibits a comparatively smaller improvement after incorporating VPCT, compared to the Joint and Reference models. This can be attributed to differences in the parameters of the Gaussian probability model employed. In the Joint and Reference models, both the mean $\mu_i$ and variance $\sigma_i$ are estimated, whereas in the Hyperprior model, only $\sigma_i$ is estimated, with $\mu_i$ always set to zero. To better understand the influence of Gaussian parameters on VPCT performance, we conducted three additional experiments on the Hyperprior, Joint, and Reference entropy models, as shown in Fig.~\ref{fig:exp:ana_gaussion_params}. We excluded the Factorized entropy model, as it is not based on a Gaussian distribution. We performed three experiments for each entropy model: 1) the entropy model without VPCT (denoted as name-vp-w/o-vpct); 2) the entropy model with VPCT using Gaussian parameter $\sigma_i$ (denoted as name-vp-w-vpct-$\sigma_i$); and 3) the entropy model with VPCT using both $\sigma_i$ and $\mu_i$ Gaussian parameters (denoted as name-vp-w-vpct-$\sigma_i$-$\mu_i$). Our experiment results revealed similar performance trends across all three entropy models in each experimental setup. Specifically, the entropy models with VPCT using both $\sigma_i$ and $\mu_i$ Gaussian parameters outperformed those with VPCT using only $\sigma_i$, which then outperformed the models without VPCT. We believe this is because VPCT's global prior information enhances the estimation of the overall trend or mean of image data by providing insights into the entire image. The variance parameter $\sigma_i$, which measures data variability and spread, is less affected by this global information as it pertains to local differences within viewports that global priors do not significantly alter. This explains why integrating the Hyperprior model with VPCT results in a significantly smaller performance improvement compared to integrating the Joint and Reference models with VPCT.

\begin{figure}[t]
  \centering
  \small
  
  \begin{minipage}{0.3\textwidth}
    \centering \textbf{GroundTruth}
  \end{minipage}%
  \begin{minipage}{0.3\textwidth}
    \centering \textbf{Reference+VPCT}
  \end{minipage}%
  \begin{minipage}{0.3\textwidth}
    \centering \textbf{Reference}
  \end{minipage}

  \vspace{1.5mm} 
  
  \begin{minipage}[t]{0.3\textwidth}
    \centering
    \includegraphics[width=\linewidth]{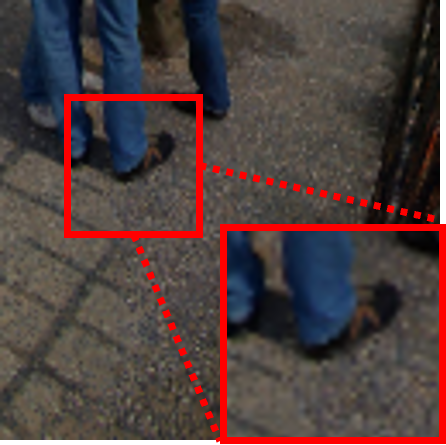}\\
    \footnotesize BPI/VPSNR
  \end{minipage}%
  \begin{minipage}[t]{0.3\textwidth}
    \centering
    \includegraphics[width=\linewidth]{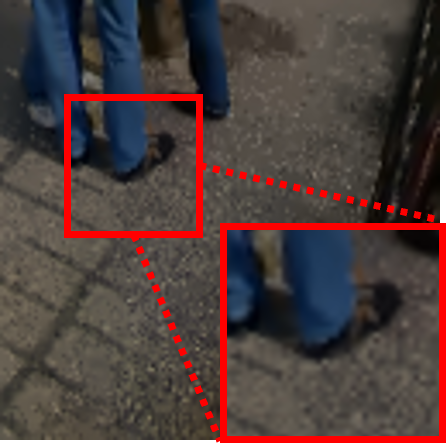}\\
    \footnotesize 0.91 (\textcolor{blue}{-2.1\%}) KB/32.85 (\textcolor{red}{+2.9\%}) dB
  \end{minipage}%
  \begin{minipage}[t]{0.3\textwidth}
    \centering
    \includegraphics[width=\linewidth]{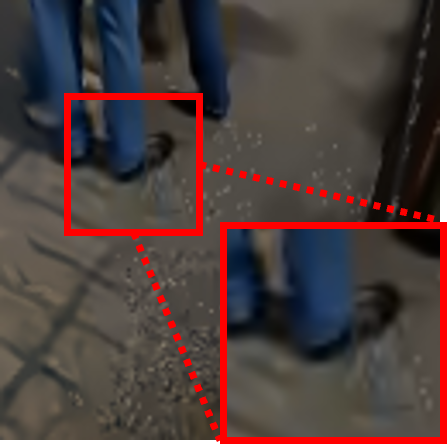}\\
    \footnotesize 0.93 KB/31.93 dB
  \end{minipage}

  \vspace{3mm}
  
  \begin{minipage}[t]{0.3\textwidth}
    \centering
    \includegraphics[width=\linewidth]{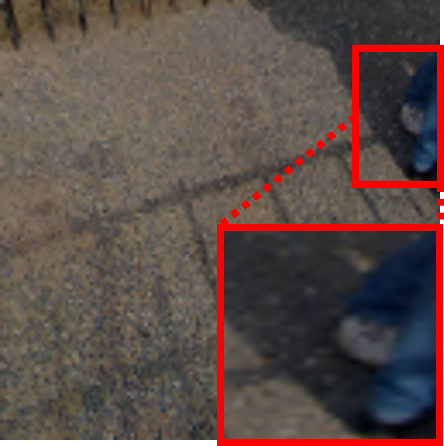}\\
    \footnotesize BPI/VPSNR
  \end{minipage}%
  \begin{minipage}[t]{0.3\textwidth}
    \centering
    \includegraphics[width=\linewidth]{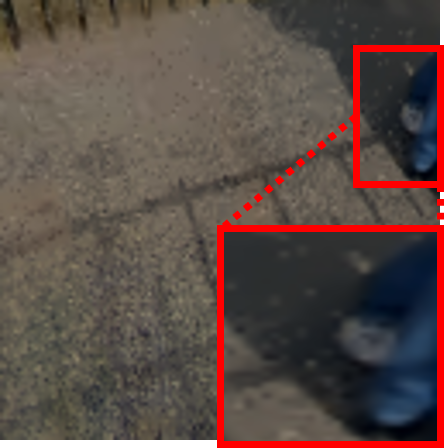}\\
    \footnotesize 0.86 (\textcolor{blue}{-3.3\%}) KB/32.65 (\textcolor{red}{+3.1\%}) dB
  \end{minipage}%
  \begin{minipage}[t]{0.3\textwidth}
    \centering
    \includegraphics[width=\linewidth]{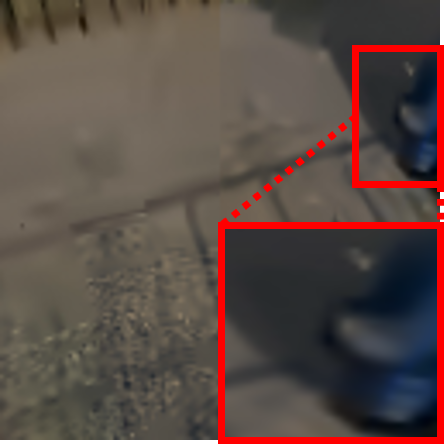}\\
    \footnotesize 0.89 KB/31.64 dB
  \end{minipage}

  \vspace{3mm}

  \begin{minipage}[t]{0.3\textwidth}
    \centering
    \includegraphics[width=\linewidth]{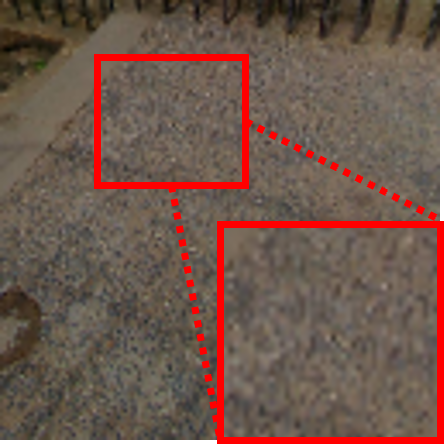}\\
    \footnotesize BPI/VPSNR
  \end{minipage}%
  \begin{minipage}[t]{0.3\textwidth}
    \centering
    \includegraphics[width=\linewidth]{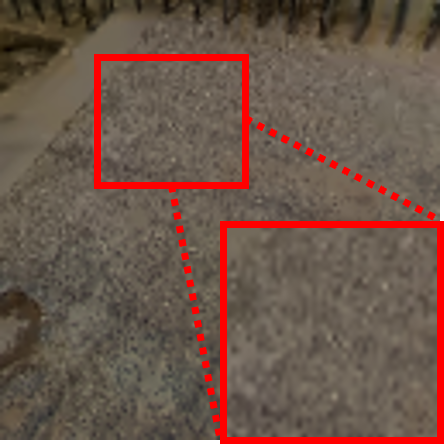}\\
    \footnotesize 0.76 (\textcolor{blue}{-3.9\%}) KB/32.22 (\textcolor{red}{+3.5\%}) dB
  \end{minipage}%
  \begin{minipage}[t]{0.3\textwidth}
    \centering
    \includegraphics[width=\linewidth]{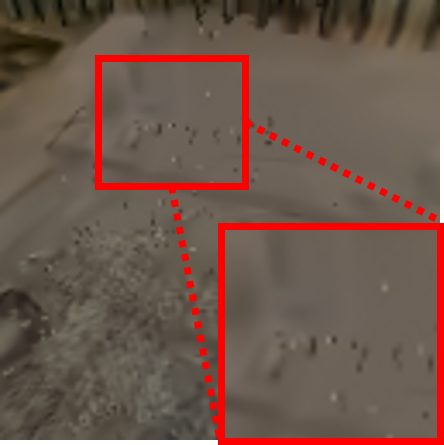}\\
    \footnotesize 0.79 KB/31.13 dB
  \end{minipage}

  \caption{Visualization of individual viewport compression.}
  \label{fig:exp:visual}
\end{figure}

\subsection{Visualization Results}

To demonstrate the benefits of VPCT in a more intuitive way, we visualize sample results in this section. 

\noindent\textbf{Viewport Reconstruction Visualization Results.}
In Fig.~\ref{fig:exp:visual}, we visualize three viewports reconstructed by the Reference model with VPCT and without VPCT as well as the raw input viewports. These viewports are compressed in the order from top to bottom. We observe that the model incorporating VPCT reconstructs viewports with higher quality and fewer bits. 
It is also worth noting that the gains for a viewport compressed later are greater than for a viewport compressed earlier. This is because the later-compressed viewport can leverage prior information from the earlier viewports, resulting in more effective compression. For example, a person wearing blue pants appears in the top viewport and this person's leg appears in the middle viewport due to the overlaps. The VPCT can utilize such previous viewport information to enhance the compression.

\begin{figure*}
\begin{tabular}{ccccc}

    \raisebox{3.ex}{\rotatebox{90}{\parbox{1.9cm}{\centering\textbf{Input\\Viewports}}}} &
    \includegraphics[width=0.19\textwidth, keepaspectratio]{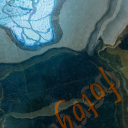} &
    \includegraphics[width=0.19\textwidth, keepaspectratio]{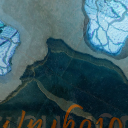} &
    \includegraphics[width=0.19\textwidth, keepaspectratio]{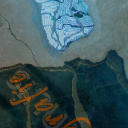} &
    \includegraphics[width=0.19\textwidth, keepaspectratio]{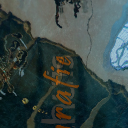} \\

    \raisebox{0.ex}{\rotatebox{90}{\parbox{2.5cm}{\centering\textbf{Redundancy \\ Map W/O VPCT}}}} &
    \includegraphics[width=0.19\textwidth, keepaspectratio]{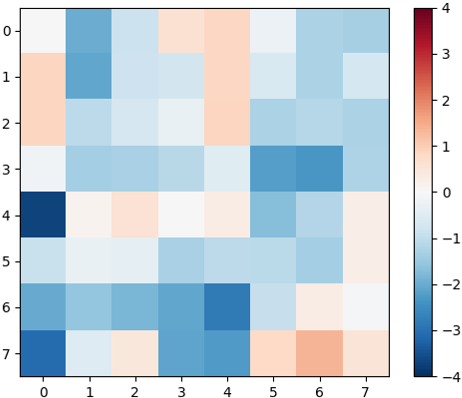} &
    \includegraphics[width=0.19\textwidth, keepaspectratio]{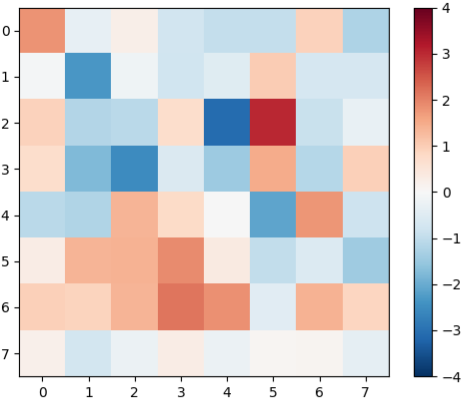} &
    \includegraphics[width=0.19\textwidth, keepaspectratio]{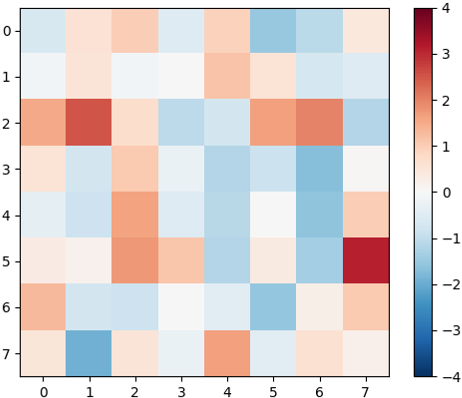} &
    \includegraphics[width=0.19\textwidth, keepaspectratio]{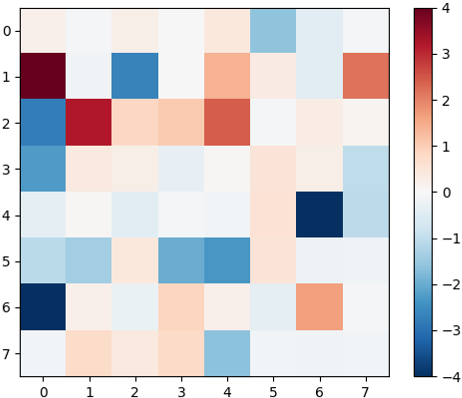} \\

    \raisebox{0.ex}{\rotatebox{90}{\parbox{2.5cm}{\centering\textbf{Redundancy \\ Map With VPCT}}}} &
    \includegraphics[width=0.19\textwidth, keepaspectratio]{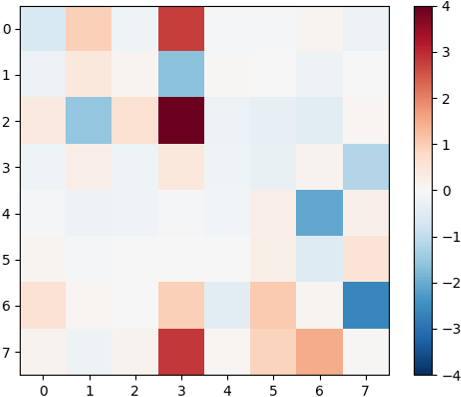} &
    \includegraphics[width=0.19\textwidth, keepaspectratio]{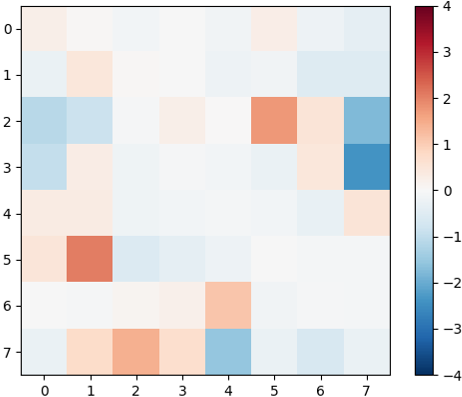} &
    \includegraphics[width=0.19\textwidth, keepaspectratio]{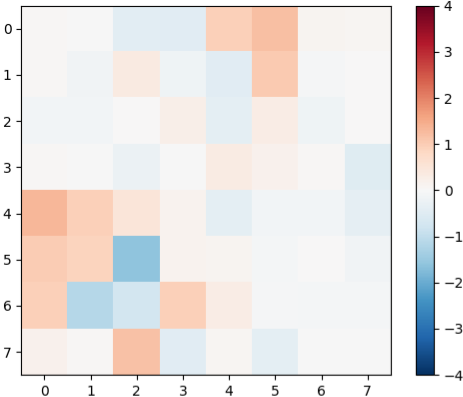} &
    \includegraphics[width=0.19\textwidth, keepaspectratio]{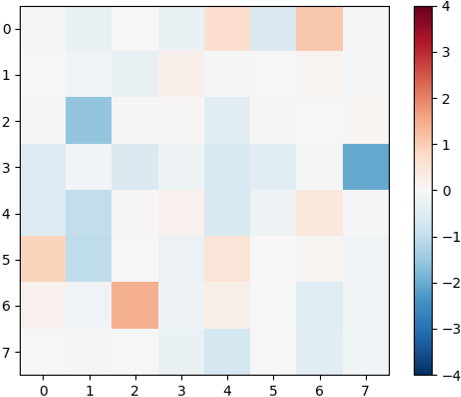}
\end{tabular}
\caption{{Visualization of the redundancy map of latent representation.}}
\label{fig:exp:visual_latent}
\includegraphics[width=.8\textwidth]{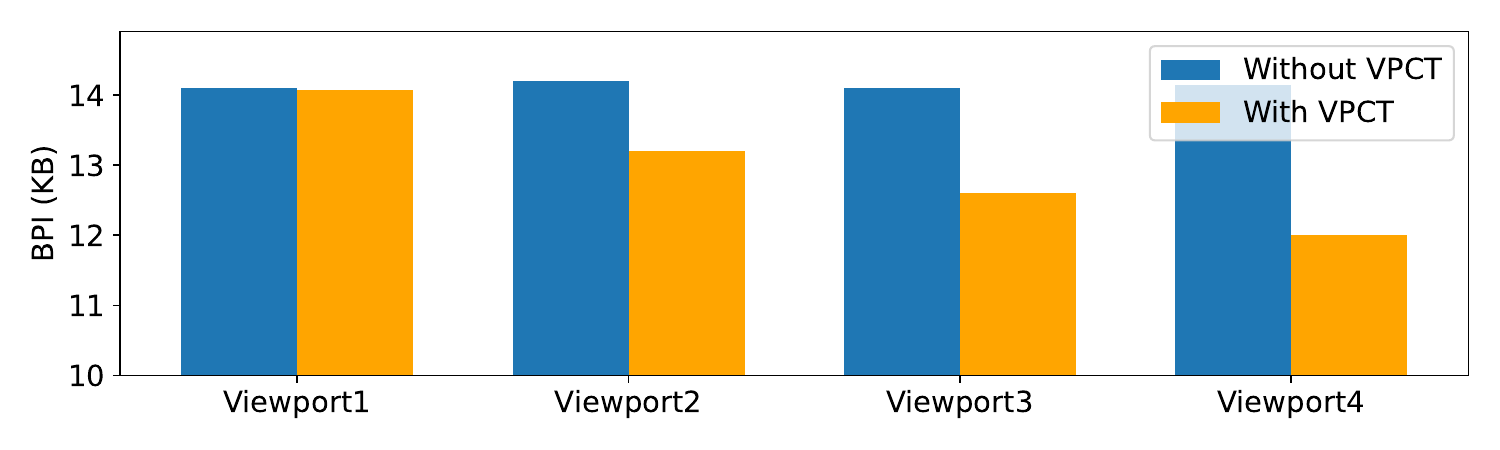}
\caption{{BPI comparison across viewports.}}
\label{fig:exp:comp_across_vp}
\end{figure*}

\noindent\textbf{Latent Redundancy Map Visualization Result.} 
{Fig.~\ref{fig:exp:visual_latent} displays the visualization results of the redundancy in the latent representation. The first row shows the input raw viewport images, arranged from left to right. The second and third rows depict the spatial redundancy maps of the entropy model without and with VPCT, respectively. The redundancy map of the latent representation is calculated as $\frac{y_i - \mu_i }{ \sigma_i^2 }$. A value closer to 0 (light colors) indicates a more accurate prediction by the entropy model, leading to higher compression efficiency. Meanwhile, we also report the per-viewport bitrate in Fig.~\ref{fig:exp:comp_across_vp}. We find that Viewport 1 shows negligible bitrate change, while Viewports 2, 3, and 4 exhibit consistently lower bitrates than the baseline without VPCT. This is because, under the sequential encoding order, Viewport 1 cannot leverage information from subsequent viewports, while Viewports 2, 3, and 4 can benefit from the contextual information provided by VPCT. Specifically, Viewports 2, 3, and 4 show bitrate reductions of 7.04\%, 10.70\%, and 15.12\% respectively, leading to an average bitrate reduction of 8.21\%. These results suggest that the VPCT module helps each viewport to remove overlap and leverage global information based on earlier viewports.}

\subsection{Ablation Studies}
\label{sec:exp:abl}
To validate the design of viewport extraction and the VPCT-based viewport codec, we conducted various ablation studies of our pipeline. 
\begin{figure}
    \centering
    \begin{minipage}[c]{\linewidth}
    \centering
    \subfigure[Impacts of extraction FoV]{
    \includegraphics[width=0.44\linewidth]{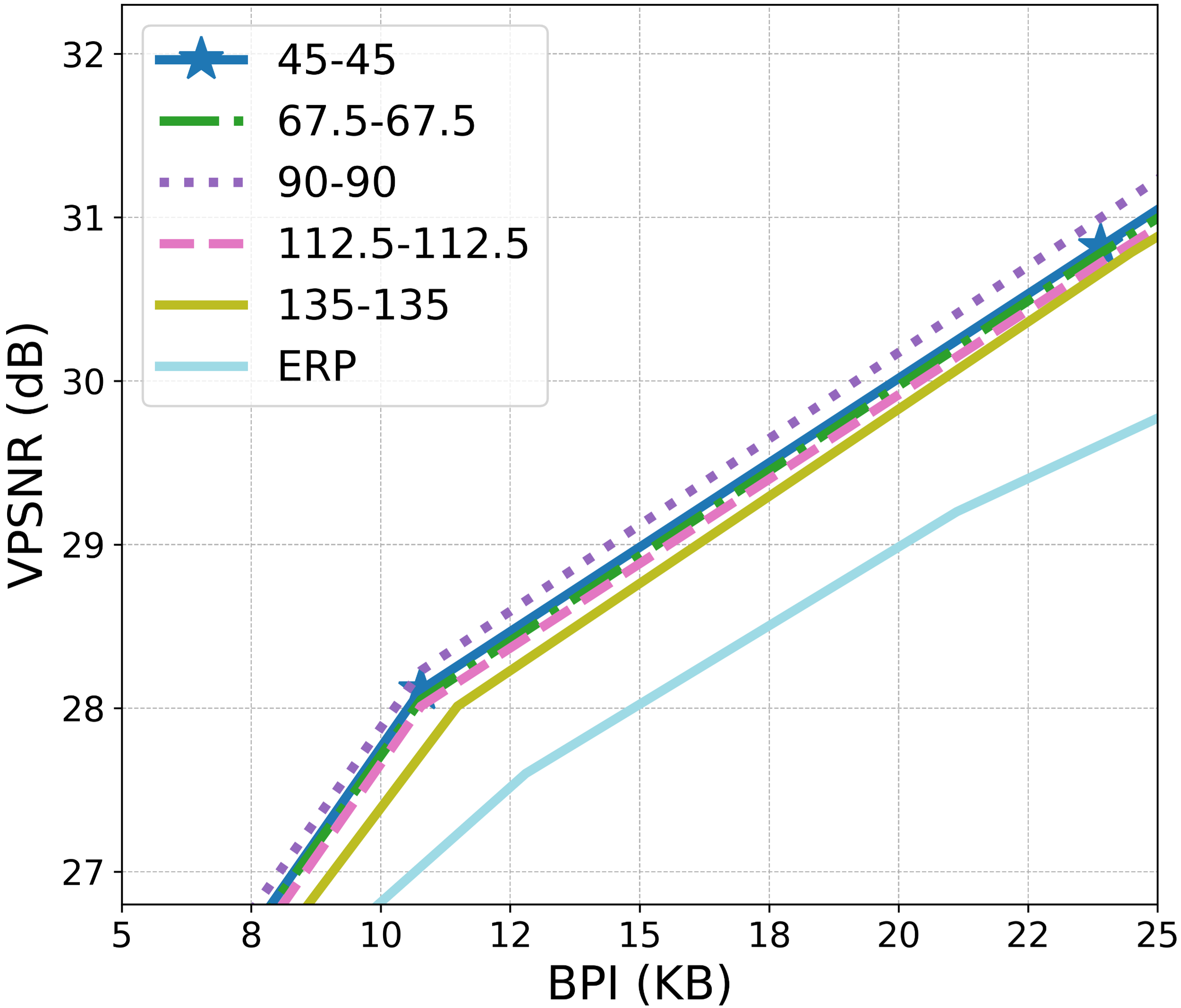}
    \label{fig:exp:vp_setting}
    }
    \subfigure[Impacts of model structure]{
    \includegraphics[width=0.44\linewidth]{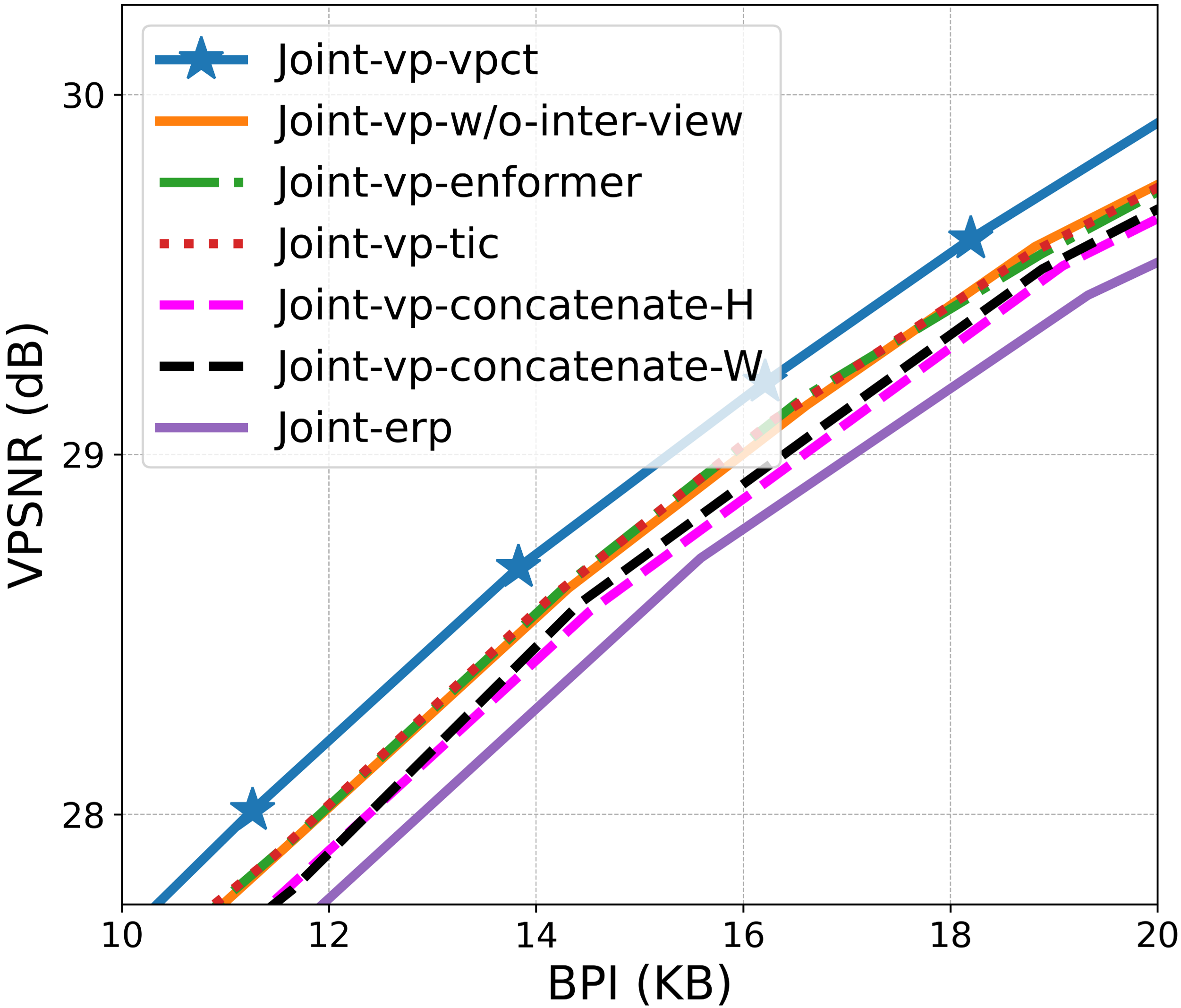}
    \label{fig:exp:abl_structure}
    }
    \end{minipage}
    \caption{Ablation studies.}
    \label{fig:exp:abl}
\end{figure}

\subsubsection{Analysis of Viewport Extraction Setting} We first explore the impact of viewport extraction settings on compression performance. The pipeline is evaluated on the Reference model
~\cite{reference}. 
As shown in Fig.~\ref{fig:exp:vp_setting}, we vary the horizontal FoV $F_h$ and vertical FoV $F_v$ concurrently from 45\degree~to 135\degree, e.g., `45-45' corresponds to a $F_h$ and $F_v$ of 45\degree~each.
Note that this setting also determines the number of extracted viewports, since different FoV covers different size of the viewports.  During the viewport extraction procedure, we start at 0\degree~latitude position, and then move the viewport extraction position northward and southward in steps of $F_v$ until reaching the positions of 90\degree~and -90\degree~latitude. For any viewport extraction latitude $\phi_C$, the number of extracted viewports equals $\lceil\frac{360\degree}{F_h}\times (\frac{C_{\phi_C}}{C_0} )\rceil$, where $C_0$ and $C_{\phi_C}$ respectively represent the circumference lengths at latitudes 0\degree~and $\phi_C$. For example, given a viewport with $F_h$ and $F_v$ both equal to 90\degree, we extract 6 viewports from the spherical image.

We observe that the viewport-based neural compression pipeline consistently outperforms the conventional pipeline using ERP across different viewport settings, which validates the robustness of the proposed pipeline. In addition, the compression performance gradually improves as the FoV decreases from 135\degree~to 90\degree. This is because a smaller viewport is more similar to the spherical area to extract, thereby mitigating the distortion and oversampling issues. However, a smaller viewport could also introduce more overlaps between viewports and lose more global information about the sphere, compromising part of its benefits, as observed with the cases of `45-45' and `67.5-67.5'.


\subsubsection{Analysis of Viewport-based Neural Compression Pipeline} This study aims to evaluate how the proposed pipeline can effectively share global prior information with the viewport codec by providing additional information to the entropy model. The proposed pipeline is evaluated on the Joint model~\cite{joint}. For comparison, we employ two alternative methods for sharing global prior information with the viewport codec. Specifically, after extracting viewports from a spherical image, we concatenate these viewports directly by height or width. Consequently, the 2D image compression model obtains global prior information by accepting concatenated viewports as input. As illustrated in Fig.~\ref{fig:exp:abl_structure}, although these two methods, denoted by Joint-vp-concatenate-H and Joint-vp-concatenate-W, improve the conventional pipeline using ERP, their performance improvement is limited in comparison to using VPCT for sharing global prior information. This is because concatenation between viewports leads to overlapping content that degrades the compression performance.

\subsubsection{Analysis of VPCT Design} The VPCT encompasses two attention blocks, the intra-view block and the inter-view block. We examine the VPCT module without the inter-view block, i.e., Joint-vp-w/o-inter-view, and implement several other transformer-based entropy models as alternatives to the VPCT for extracting global prior information, including \mbox{Joint-vp-tic}~\cite{tic} and Joint-vp-enformer~\cite{entroformer}. The results in Fig.~\ref{fig:exp:abl_structure} show that these three alternative approaches achieve comparable results, but they all exhibit notably inferior performance compared to the proposed approach with the full VPCT module. One explanation for the limitations of existing transformer-based entropy models is their inability to efficiently exchange information across different viewports and address the redundant information in the overlapping viewports.

\subsection{Complexity and Latency}

{
To assess the computation complexity, latency, and memory usage, we measured the Floating Point Operations (FLOPs), MAC Per Pixel (MPP), encoding/decoding time, and GPU memory consumption of the proposed pipeline. }
We used the Reference model~\cite{reference} as the underlying 2D entropy model as a sample case to evaluate the computation cost.

\noindent
\textbf{Compression Complexity.} 
To evaluate the computational complexity of the proposed pipeline, we measured the FLOPs and MPP, with and without the VPCT module. The results are presented in Table~\ref{tab:complex2}. Our findings indicate that the VPCT module introduces only a minimal increase in computational complexity. Specifically, the VPCT adds 2.30\% more FLOPs and 2.32\% more MPP to existing 2D codecs. 
The increase in computational complexity is justified by the 11.11\% improvements in compression performance of the proposed pipeline over the default Reference model.

\noindent
\textbf{Compression Latency.} To evaluate the compression latency of the proposed pipeline, we measured the encoding and decoding time. We also evaluate an accelerated implementation using Checkerboard~\cite{checkerboard}, which is a parallel algorithm designed for autoregressive entropy models that reduces the computational resources of 2D entropy models. Since the proposed viewport-based pipeline is compatible with canonical 2D entry models, we can utilize this technique to optimize the computation speeds. Our findings in Table~\ref{tab:complex} indicate that the Reference model experiences a 4.7$\%$ and 5.8$\%$ increase in encoding and decoding time after including the VPCT. However, thanks to the Checkerboard technique, we can roughly reduce the computation time by half, making the codec even faster than the SOTA 360\degree image compression model~\cite{structure_map}. 

\noindent{\textbf{Compression GPU Memory Usage.} To compare the GPU memory usage of the proposed pipeline with baseline methods, we measured the peak GPU memory consumption during both encoding and decoding processes. As shown in Tab.~\ref{tab:memory}, replacing the ERP-based pipeline with a viewport-based pipeline significantly reduces GPU memory usage during both encoding and decoding. Specifically, switching from the ERP-based pipeline (denoted as Reference-erp) to the viewport-based pipeline (denoted as Reference-vp) decreases GPU memory consumption by 33.7\% and 44.1\% during encoding and decoding, respectively. This reduction is mainly because the viewport-based pipeline processes multiple smaller viewport images instead of a single large ERP image, thereby lowering memory usage. Although the VPCT module introduces a slight memory overhead of 1.5\% and 1.3\% during encoding and decoding, respectively, the overall GPU memory usage of Reference-vp (with VPCT) remains significantly lower than that of Reference-erp in both stages.
}

\begin{table}[]
    \centering
    \renewcommand{\arraystretch}{1.2}
    \resizebox{\textwidth}{!}{
    \begin{tabular}{m{6cm} cc}
        \toprule
        Codecs & FLOPs (G) & MAC Per Pixel (K) \\ \hline\hline
        Reference-vp & 176.53 & 457.28 \\ 
        Reference-vp (with VPCT) & 180.59 \textcolor{blue}{(+2.30$\%$)} & 468.15 \textcolor{blue}{(+2.32$\%$)} \\ \bottomrule
    \end{tabular}
    }
    \caption{The increased amount of model calculations.}
    \label{tab:complex2}
\end{table}


\begin{table}[h]
    \centering
    \renewcommand{\arraystretch}{1.2}
    \resizebox{\textwidth}{!}{
    \begin{tabular}{m{8cm} cc}
        \toprule
        \multirow{2}{*}{Codecs} & Encoding & Decoding \\
                                & Time (s) & Time (s) \\ \midrule \midrule
        SOTA~\cite{structure_map} & 3.07 & 3.50 \\ 
        Reference-erp~\cite{reference} & 3.40 & 5.14 \\
        Reference-vp-w-VPCT & 3.56 \textcolor{blue}{(+4.7\%)} & 5.44 \textcolor{blue}{(+5.8\%)} \\
        Reference-vp-w-VPCT-Checkerboard & 1.82 \textcolor{red}{(-46\%)} & 2.32 \textcolor{red}{(-55\%)} \\
        \bottomrule
    \end{tabular}
    }
    \caption{Comparison of encoding and decoding time.}
    \label{tab:complex}
\end{table}

\begin{table}[h]
    \centering
    \renewcommand{\arraystretch}{1.2}
    \resizebox{\textwidth}{!}{
    \begin{tabular}{m{7cm} cc}
        \toprule
        \multirow{2}{*}{{Codecs}} & Encoding & Decoding \\
                                         & Memory (MB) & Memory (MB) \\ \midrule \midrule
        Reference-erp & 1488.00 & 1752.00 \\
        Reference-vp  & 986.00 (\textcolor{red}{-33.7$\%$}) & 980.00 (\textcolor{red}{-44.1$\%$}) \\ 
        Reference-vp (with VPCT) & 1008.67 (\textcolor{red}{-32.2$\%$}) & 1002.70 (\textcolor{red}{-42.8$\%$}) \\ 
        \bottomrule
    \end{tabular}
    }
    \caption{{GPU-memory usage comparison}}
    \label{tab:memory}
\end{table}

\section{Conclusion and Future Work}
We propose the first viewport-based neural compression pipeline for 360\degree~images to minimize the well-known oversampling and distortion issues. To address the content overlaps and information redundancy between extracted viewports, we design a novel VPCT module for maximal viewport compression. Furthermore, the proposed pipeline can be seamlessly integrated with canonical learning-based 2D image compression models. Experimental results show that our pipeline achieves state-of-the-art performance compared to standard codecs and learning-based 360\degree~image codecs and 2D image codecs. The performance gain is consistent across various underlying entry models and model design factors. Despite the strong results, our pipeline has several limitations. Its performance depends on the viewport FoV---the gain from cross-viewport context modeling may decrease under extreme FoV settings. In addition, our method cannot reduce the bitrate of the first viewport due to the sequential encoding order, leading to uneven bitrate reduction across viewports.
{
To overcome these challenges, future research should focus on developing adaptive FoV strategies that dynamically adjust viewport configurations based on content characteristics. This could improve robustness across different FoV settings and enhance cross-viewport context modeling. Furthermore, to address the limitation of sequential encoding order, one could explore bi-directional or hierarchical prediction structures to distribute bitrate savings more uniformly across viewports.} 

\bibliographystyle{ACM-Reference-Format}
\bibliography{sample-base}

\end{document}